\RequirePackage{fix-cm}
\documentclass[smallextended]{svjour3}       % onecolumn (second format)
\smartqed  % flush right qed marks, e.g. at end of proof
\usepackage{graphicx}
\begin{document}

\title{PACS photometer calibration block analysis}



\author{Mo\'or, A.  \and
        M\"uller, Th.~G. \and
	Kiss, Cs. \and 
	Balog, Z. \and
	Billot, N. \and
	Marton, G. 
}


\institute{A. Mo\'or \at
              Konkoly Observatory, MTA CSFK \\
              Tel.: +36-1-3919341\\
              Fax: +36-1-2754668\\
              \email{moor@konkoly.hu}           %  \\
           \and
           Th. M\"uller \at
           Max-Planck-Institut f\"ur extraterrestrische Physik, Garching, 
	   Germany
	   \and 
	   Cs. Kiss and G. Marton \at
	   Konkoly Observatory, MTA CSFK
	   \and
	   Z. Balog \at
	   Max-Planck-Institut f\"ur Astronomie, Heidelberg, Germany
	   \and 
	   N. Billot \at
	   Instituto de Radio Astronom\'\i a Milim\'etrica, Granada, Spain      
}

\date{Received: date / Accepted: date}

\maketitle

\begin{abstract}
The absolute stability of the PACS bolometer response over the entire mission
lifetime without applying any corrections is about 0.5\% (standard deviation) or
about 8\% peak-to-peak. This fantastic stability allows us to calibrate all
scientific measurements by a fixed and time-independent response file,
without using any information from the PACS internal
calibration sources. However, the analysis of calibration block observations
revealed clear correlations of the internal source signals with the evaporator
temperature and a signal drift during the first half hour after the cooler
recycling. These effects are small, but can be seen in repeated measurements
of standard stars. 
From our analysis we established corrections for
both effects which push the stability of the PACS bolometer response to about
0.2\% (stdev) or 2\% in the blue, 3\% in the  green and 5\% in the red channel
(peak-to-peak). After both corrections we still see a correlation of the 
signals with PACS FPU temperatures, possibly caused by parasitic
 heat influences via the Kevlar wires which connect the
bolometers with the PACS Focal Plane Unit.
No aging effect or degradation of the photometric system during the mission 
lifetime has been found.

%Insert your abstract here. Include keywords, PACS and mathematical
%subject classification numbers as needed.
\end{abstract}

\section{Introduction}
\label{intro}
During the nearly four-year long mission of the {\sl Herschel Space 
Observatory} (Pilbratt et al., 2010) more than 21000 photometric observations 
were performed with the Photoconductor Array Camera \& Spectrometer (PACS, 
Poglitsch et al., 2010) onboard the spacecraft.
PACS photometric observations were performed by two bolometer arrays 
that operate parallel acquiring images at 70 or 100~$\mu$m (blue array) 
and at 160~$\mu$m (red array). {These observations, by their 
nature, were heterogeneous in terms of different characteristics, including the utilized observing mode, brightness     
of the target and the sky background, the duration of the measurement.} 
However, each PACS photometric observation includes at 
least one calibration 
block (CB) measurement that consists of chopped observations between the two PACS 
internal calibration sources (CSs). CB observations -- on the contrary to science measurements -- 
 are always performed using a standard, identical manner.
%and which performed standard way 
%during the entire mission. 
Thus, they
constitute a homogeneous data set that enables 
the monitoring of the number of dead, bad/noisy or saturated pixels and particularly 
the possible short- and long-term
  evolution of bolometer response during the 
full mission lifetime on the basis of a well-defined differential signal from
the PACS calibration sources. Calibration blocks are also very useful
  to verify changes on a given operational day (OD, day elapsed since the launch of the spacecraft) 
  or a sequence of ODs for
  science projects with repeated measurements of the same
  source in a short period of time.
 
In order to investigate the behaviour of the bolometer responsivity,
we compiled a database of photometer calibration block observations 
by processing all PACS photometric measurements, including those 
obtained in 
{SPIRE/PACS} parallel mode. 
 Using this database we searched for the  
dependencies of {detector signal output} on various instrument and satellite parameters.
{A prelimininary analysis of the CB signals and the features observed in the first year 
of the {\sl Herschel} mission are summarized in Billot et al. (2010).} 
%{\bf bolometers' signal output}

\section{Calibration block observations and data processing strategy}
\label{sec:1}

\subsection{Internal calibration sources}
\label{sec:cs}

\begin{figure*}
  \includegraphics[width=0.45\textwidth]{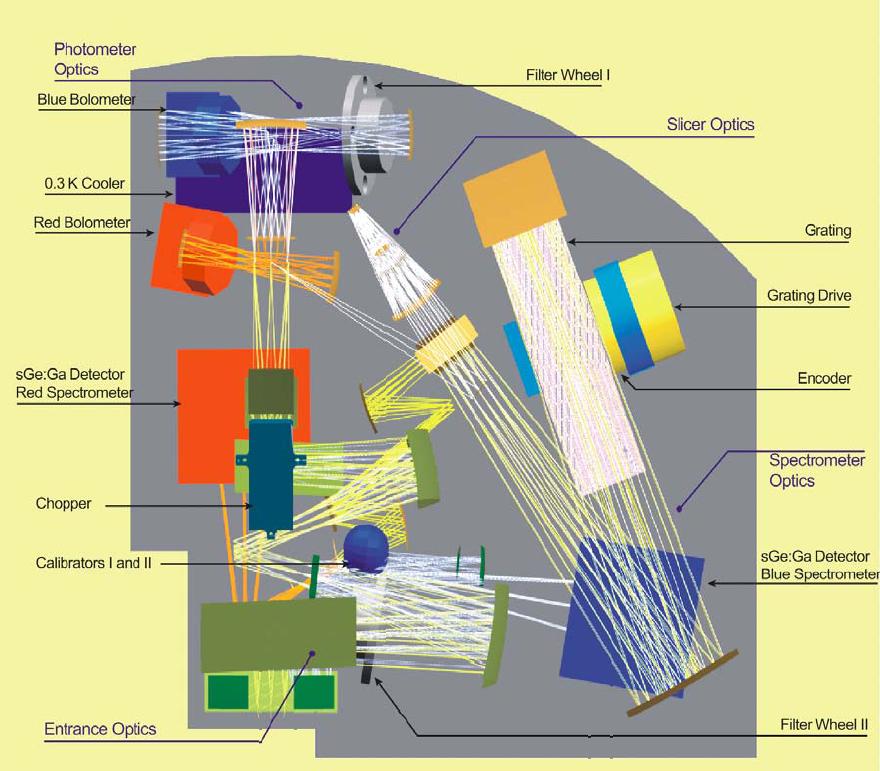}
  \includegraphics[width=0.55\textwidth]{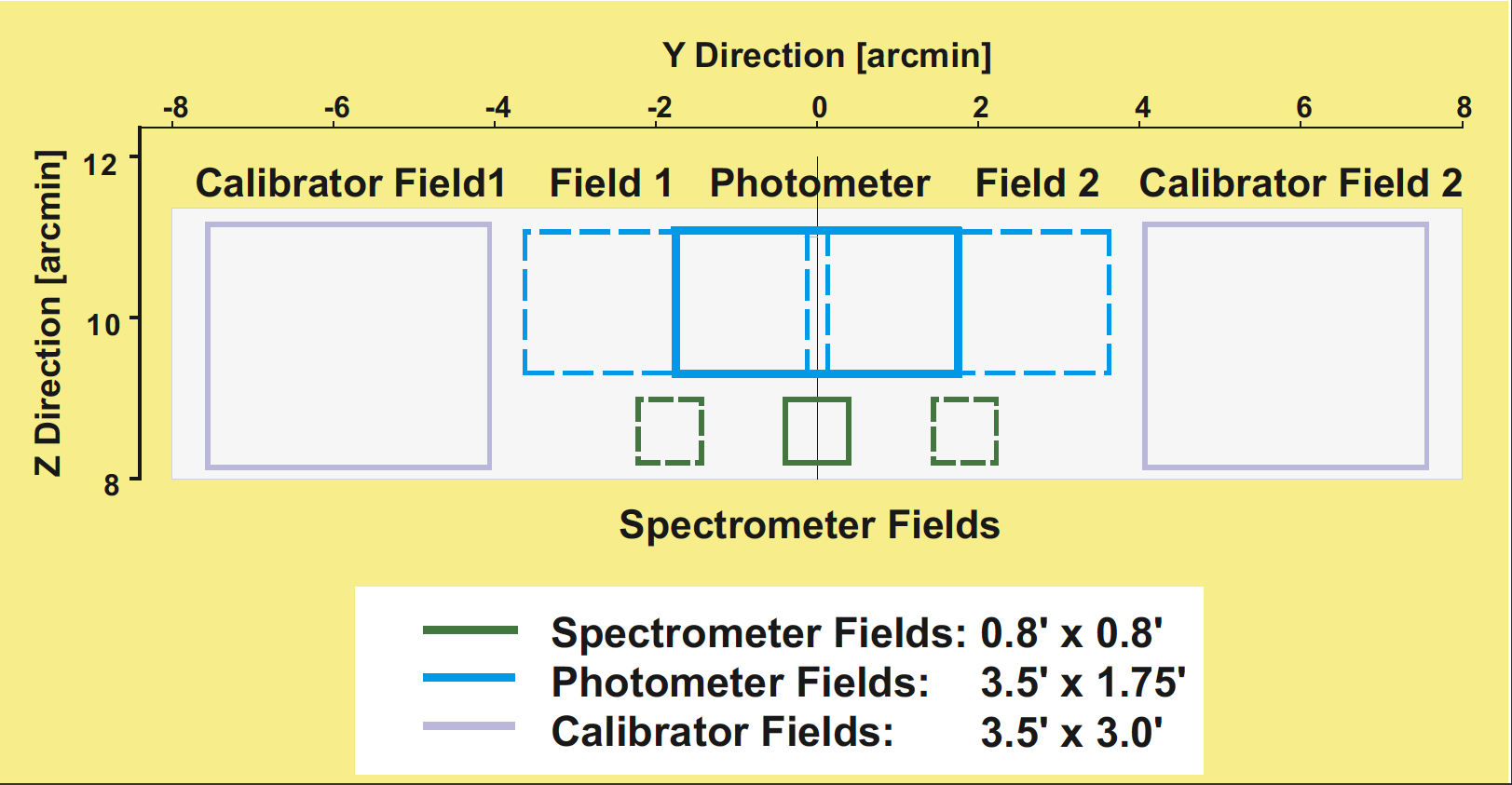}
\caption{{\sl Left: Optical layout of PACS. Calibration sources are placed at the entrance of the instrument.
Right: The PACS field-of-view in combination with the chopper positions of the optical axis and the
positions on both PACS calibration sources (CSs). In the PHOT calibration block chopping between the two
CSs is performed. Both figures are taken from the PACS Observer's Manual.}
}
\label{calsources}       % Give a unique label
\end{figure*}

For the continuous monitoring of the detector noise and responsivity 
we need a source of constant flux that is provided by  
the internal calibrators within the PACS instrument.
{The two grey body calibration sources are placed near the entrance to
the instrument (Fig.~\ref{calsources} left), outside of the Lyot stop, 
to have approximately the same
light path for observation and internal calibration.
They provide far-infrared radiation loads slightly below 
(CS1, 48~${\rm \Omega}$, 55~K) and above
(CS2, 58~${\rm \Omega}$, 60 K) the telescope background at around 100{$\mu$m}
(both loads are below the telescope background at the shortest
wavelengths and above at the longest wavelengths due to differences
between the telescope and calibration sources emissivities).
They uniformly illuminate both the field of view and the Lyot
stop, to mimic the illumination by the telescope.}

{Details about the calibration sources are specified in the "Herschel
PACS Calibration Source Interface Control Document" (PACS-KT-ID-007,
Issue 4, Kayser-Threde GmbH, April 2004). At typical operation
temperatures each source has an average power of well below 1.6~mW,
at heating up phases a peak power of 10~mW can be applied, resulting
in maximum currents of up to 11~mA at the lower temperature range
and about 7~mA at 100~K.

%Each emitter is an assembly of two PT500 resistors (Model 118 I1F,
%Goodrich), adjustable in the temperature range of 50 to 100~K. The
%heater sensor control electronics guarantees a stability of better
%than 5~mK by using a 4-wire technique. The 5~mK level corresponds
%to $<$ 5~m${\rm \Omega}$ at 50~K, requiring a 15 bit resolution for the specified
%temperature range.
The
heater sensor control electronics guarantees a stability of better
than 5~mK that corresponds
to $<$ 5~m${\rm \Omega}$ at 50~K, requiring a 15 bit resolution for the specified
temperature range.}

\begin{figure*}
%\centering
  \includegraphics[width=0.50\textwidth]{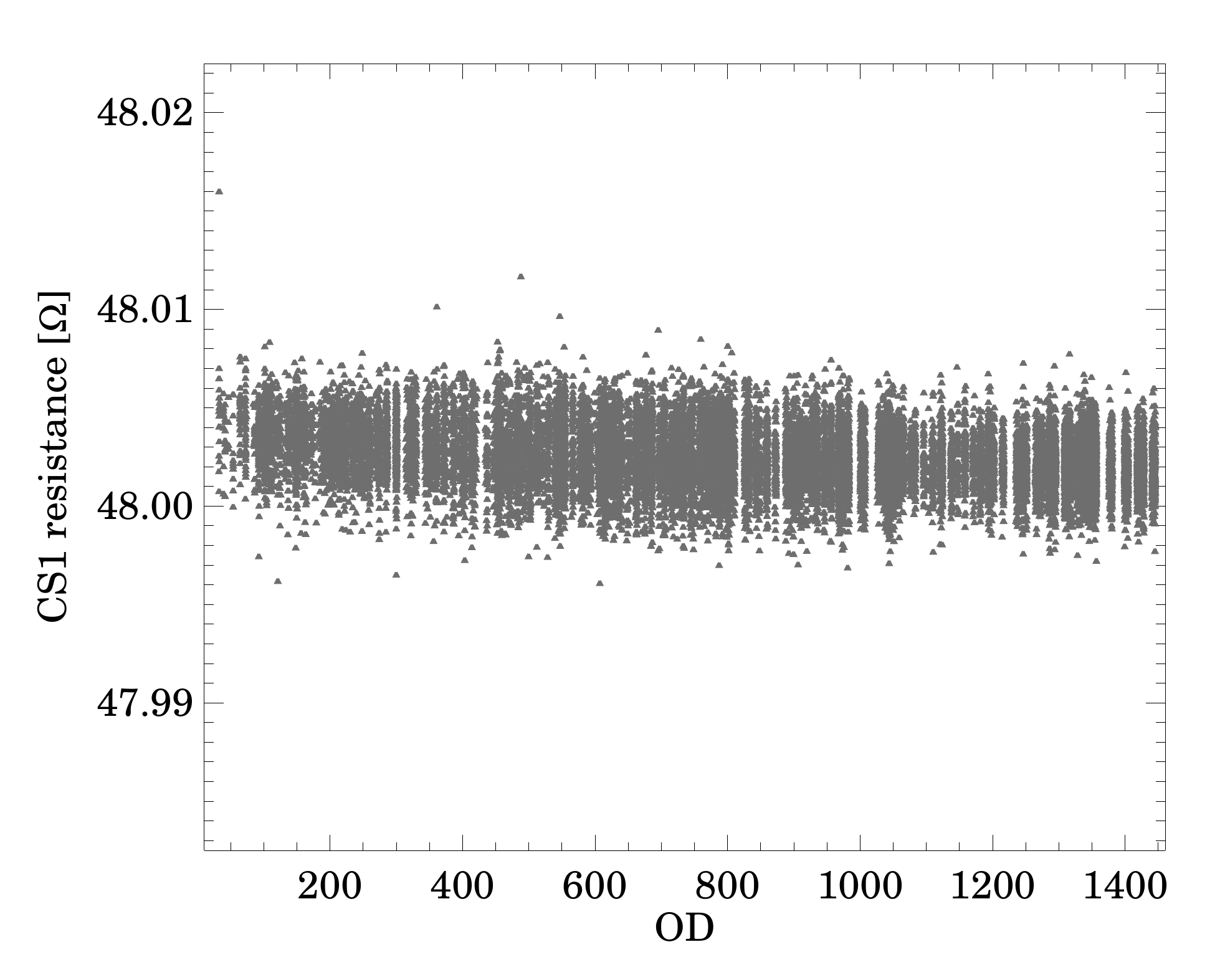}
  \includegraphics[width=0.50\textwidth]{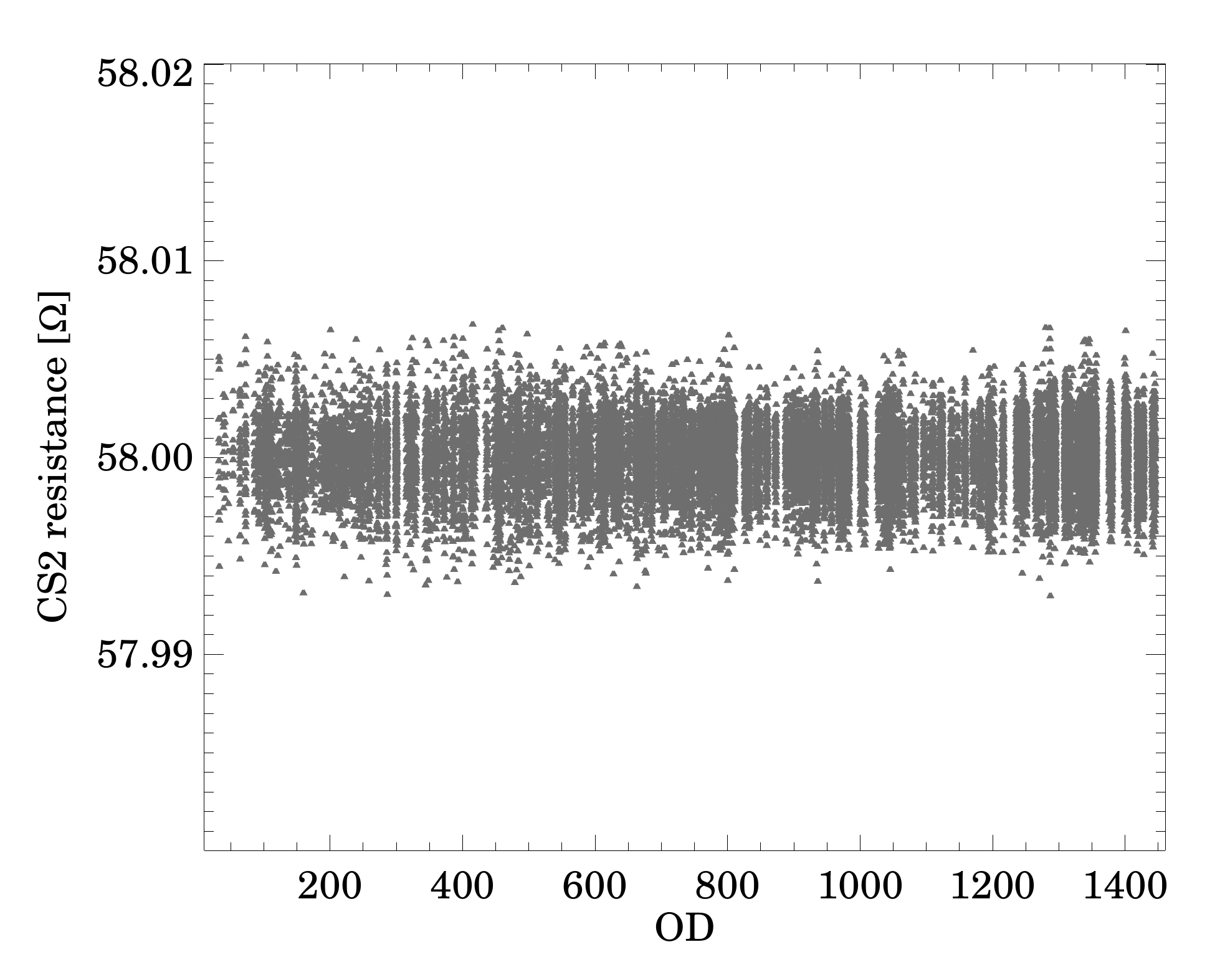}
\caption{{\sl CS1 (left) and CS2 (right) resistance values as a function of OD.}
}
\label{csres}       % Give a unique label
\end{figure*}

{The output load of the calibration sources was calibrated during
instrument level tests against two highly stable and well characterised
black bodies inside the PACS optical ground segment equipment
(OGSE black bodies). The observed resistance values (see Fig.~\ref{csres}) are
48.0004$\pm$0.0015 ${\rm \Omega}$ (13 m${\rm \Omega}$ peak-to-peak) for CS1 and
58.0000$\pm$0.0019 ${\rm \Omega}$ (14 m${\rm \Omega}$ peak-to-peak) for CS2.
The m${\rm \Omega}$ variations can be easily translated into mK temperature
variations via the calibration certificate (chapter 31 of the cryogenic
qualification model Kaiser-Threde GmbH end-item data package):
A 20 m${\rm \Omega}$ resistance variation corresponds roughly to a 10~mK
temperature variation (in the 55 to 60~K range) which is about 0.05
per cent in flux or surface brightness, well below the bolometer
measurement accuracy. Our extracted peak-to-peak resistance
values are well within the assumed 20 m${\rm \Omega}$. As expected from these
calculations, we found no correlation between the residual fluctuations
(seen after all corrections have been applied) and the measured
calibration source resistance values.}

\subsection{Design of calibration block observations}
\label{sec:cb}
A calibration block observation contains several chopper cycles on 
both PACS calibration sources (Fig.~\ref{calsources} right). The chopper moves with a frequency of 0.625~Hz 
between the two PACS CSs. In total 19 chopper cycles are executed within a 
given measurement, each chopper plateau lasts for 0.8\,s producing 8 frames in 
the down-link (4 bolometer readouts are averaged), hence a 
CB observation lasts for about 30\,s in total. 
In the {SPIRE/PACS} parallel mode the setting for blue array CB measurements is 
slightly different since 8 bolometer readouts are averaged in the blue 
channel, in the red channel 4 readouts are averaged exactly as in the PACS prime mode. 
%(instead of 4 readouts as in the case of other PACS photometric CBs). 
 
A typical PACS photometer measurement contains 
one calibration block observation, 
that precedes the science part of the measurement in the observational 
sequence. However, 
there are some observations for which more than one CB measurements were obtained, e.g. 
in the case of {SPIRE/PACS} parallel mode observations the science block are typically bracketed 
by two CBs.  
% Calibrational purpose
%Similarly to science observations the blue/red bolometers operate parallel during 
%calibration block measurements by using the same filter setup. 
%Calibration block measurements used the same filter setup as the actual science observation.
Calibration blocks are executed exactly in the same filter, gain and SPU setting as the 
science block that they belong to.
CBs are performed in the target acquisition phase. 
Up to the OD545
%545$^{\rm th}$ operational day (OD545)
%, day elapsed since the launch of the spacecraft), 
the calibration block was always executed towards 
the end of the target acquisition phase (slew) and the science part 
was separated from the calibration block by only 5~s delay time.
However, because the bolometer signals display a noticeable drift after the calibration 
block for about 30\,s this strategy was revised and after OD546 
the CBs were executed at the earliest possible
moment during the slew allowing in many cases for a longer signal stabilisation 
before the science observation started.

Due to their design, calibration blocks can provide a clean start for the subsequent
science observations and allow the determination of bolometer response using
the derived differential signal of PACS calibration sources. 	

\subsection{Data processing of CB observations}
\label{sec:dataprocessing}
\begin{figure*}
  \includegraphics[width=0.50\textwidth]{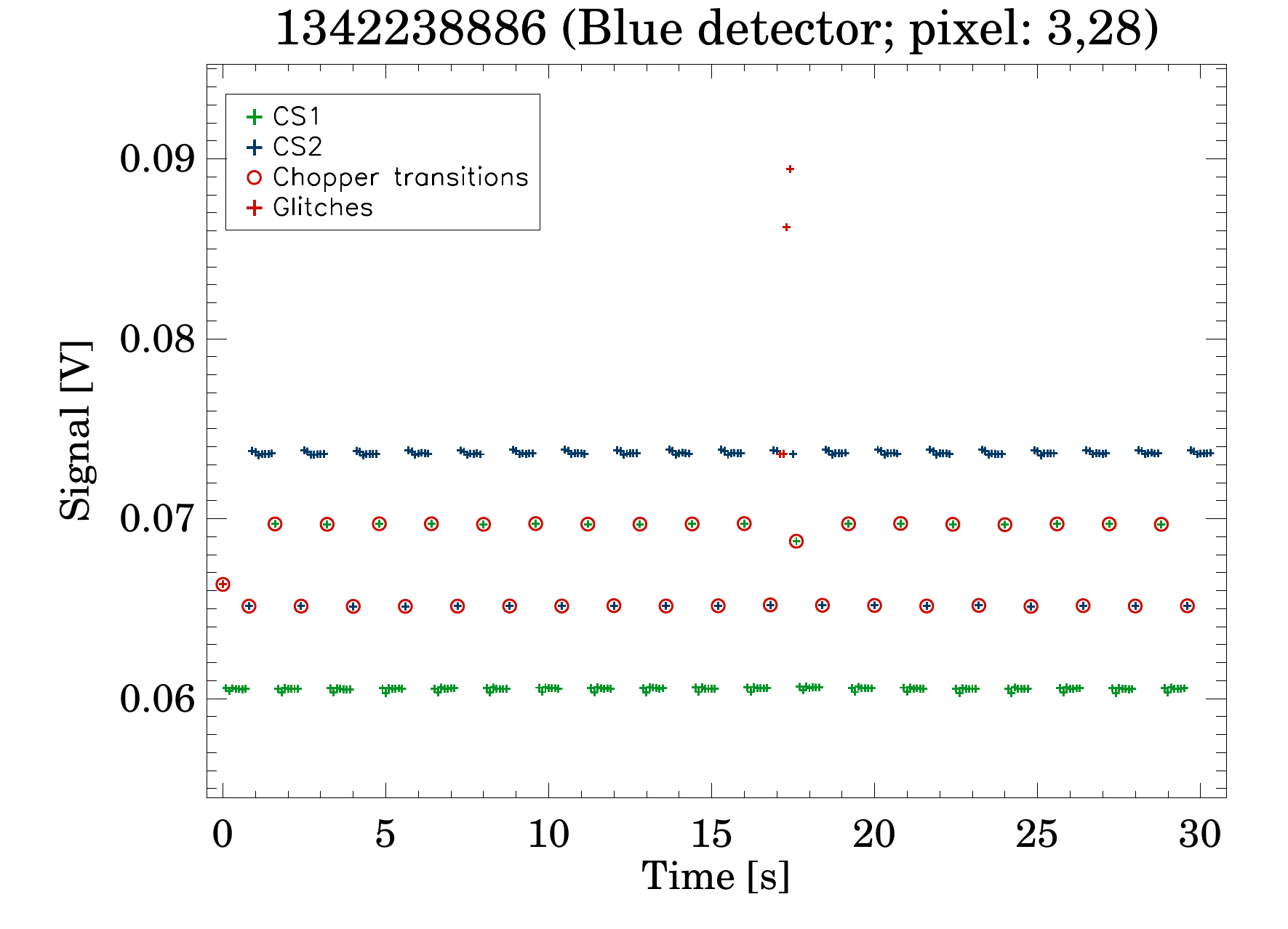}
  \includegraphics[width=0.50\textwidth]{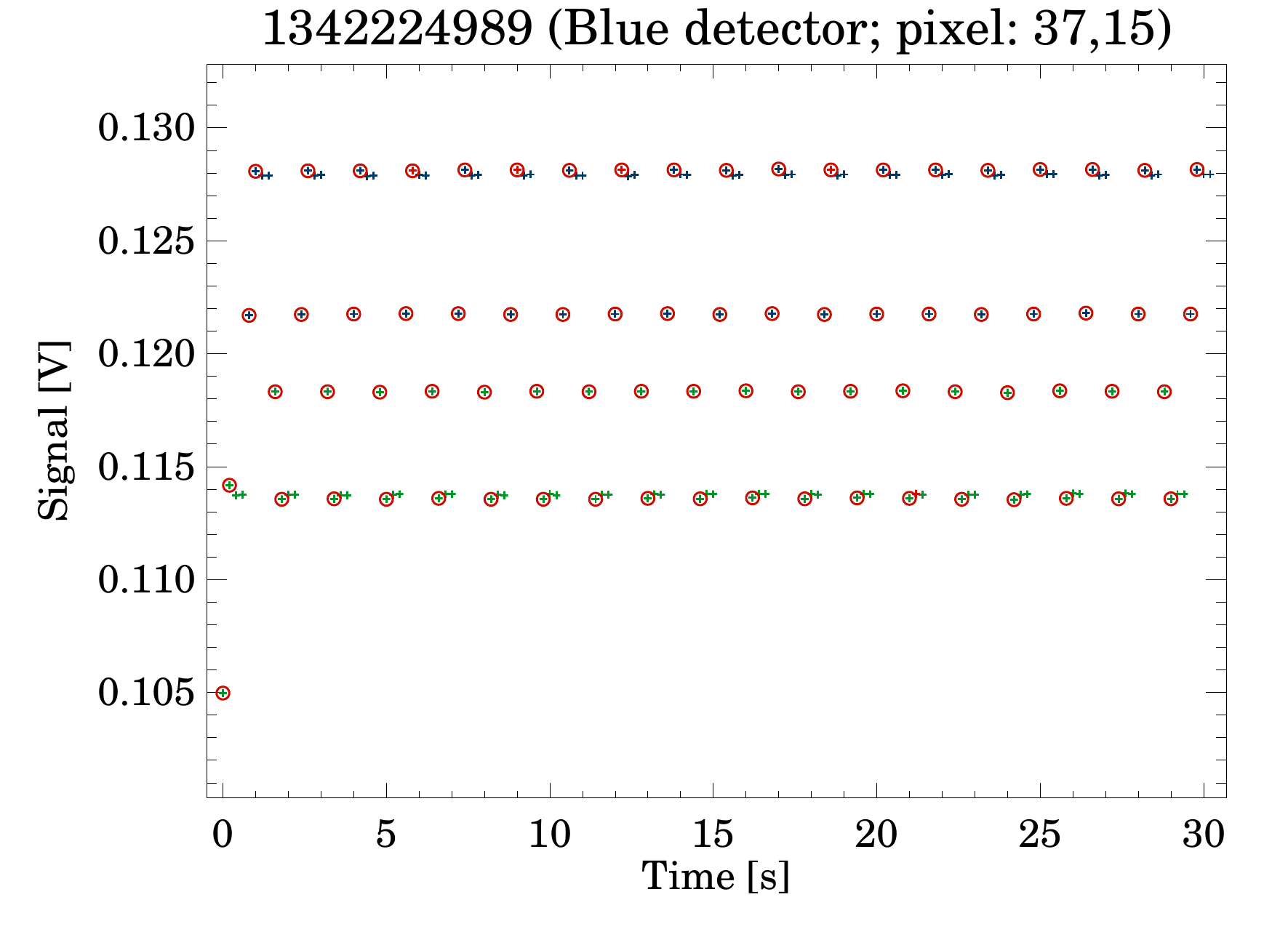}
\caption{{\sl Data stream in a typical calibration block observation obtained with the blue detector 
in the PACS prime mode (left) 
and in the PACS parallel mode (right). The two CS signals were
  derived from the cleaned chopper plateaus.}
}
\label{calblock}       % Give a unique label
\end{figure*}

Our CB processing scheme includes the following steps: 
1) identify the calibration block(s) within the specific measurement 
(perform {\sl detectCalibrationBlock} and  
 {\sl findBlocks}, then select data points that belong to the CB); 2) perform basic data 
 processing steps ({\sl photFlagBadPixels, photFlagSaturation, photConvDigit2Volts,  
   photMMTDeglitching}); 3) after eliminating chopper
  transitions (the first point in each chopper plateau for PACS prime observations and for 
  PACS parallel observations with red detector, the first two points in the case of blue 
  parallel CBs) and 
  glitches, determine the CS1 and CS2 signal levels for each pixel 
  ($S_{CS1}^{i,j}$, $S_{CS2}^{i,j}$, where $i$ and $j$ are the $i^{th}$ and $j^{th}$ pixels 
  of the blue or red array) 
  by computing {an outlier-resistant average of chopper plateaus\footnote{Applying IDL's {\sl biweight\_mean} routine that uses 
  Tukey's biweight method for robust moment estimation.}} 
  %(using IDL's biweight mean routine)}. 
  Then differential signals are determined as $S_{diff}^{i,j} = S_{CS2}^{i,j} - S_{CS1}^{i,j}$ 
  for each detector pixel.
  Figure~\ref{calblock} displays the signal in a typical calibration block observation.

As a final step we 1) calculate robustly averaged CS1, CS2 and differential signals for the 
blue and red array using all unmasked pixels, and 2) extract 
additional relevant parameters from the measurement and the 
related housekeeping data that are neccessary for the trend analysis
(e.g. the FineTime of the observation, evaporator temperature, etc.). 

\subsection{Database of CB observations}
\label{sec:database}
Altogether 20795 PACS photometric ({including 854 SPIRE/PACS parallel}) 
observations are available in the Herschel archive. This number includes 
both scientific and calibration observations.
We processed all CBs that belong to these measurements.
Although most observations include only one CB measurement, science and calibration
  observations in early mission phases (up to OD150) were performed
  with more than one (up to 18) CBs. {SPIRE/PACS} parallel mode observations typically contain 
  two CBs.
 Therefore our final database 
contains 22000 calibration block 
observations in total.

%\section{Investigation of differential CB signals dependencies on different 
%instrumental parameters}
\section{Results}
\label{sec:1}

%\subsection{Evolution of CS differential signals}
The actual bolometer response is well characterized by the differential signal of 
the two internal calibration sources measured in the corresponding CB observation.
By investigating the evolution of differential CS signals throughout the {\sl Herschel} 
mission we can study the long term behaviour of the bolometer response.
As a first step of this study we searched for possible systematic trends between 
differential signals and instrumental parameters. 
Then we developed corrections to the revealed trends and applied them 
to our data. Finally we used the corrected differential signals to study 
the long term variations of bolometer response.

We found that $S_{diff}$ correlates well both with the evaporator temperature and 
the temperature of the focal plane unit. 
Moreover, signal levels also show a short drift in the first
  0.5\,h after the cooler recycling has finished.

\subsection{Correlation with evaporator temperature}
\begin{figure*}
\centering
  \includegraphics[width=0.70\textwidth]{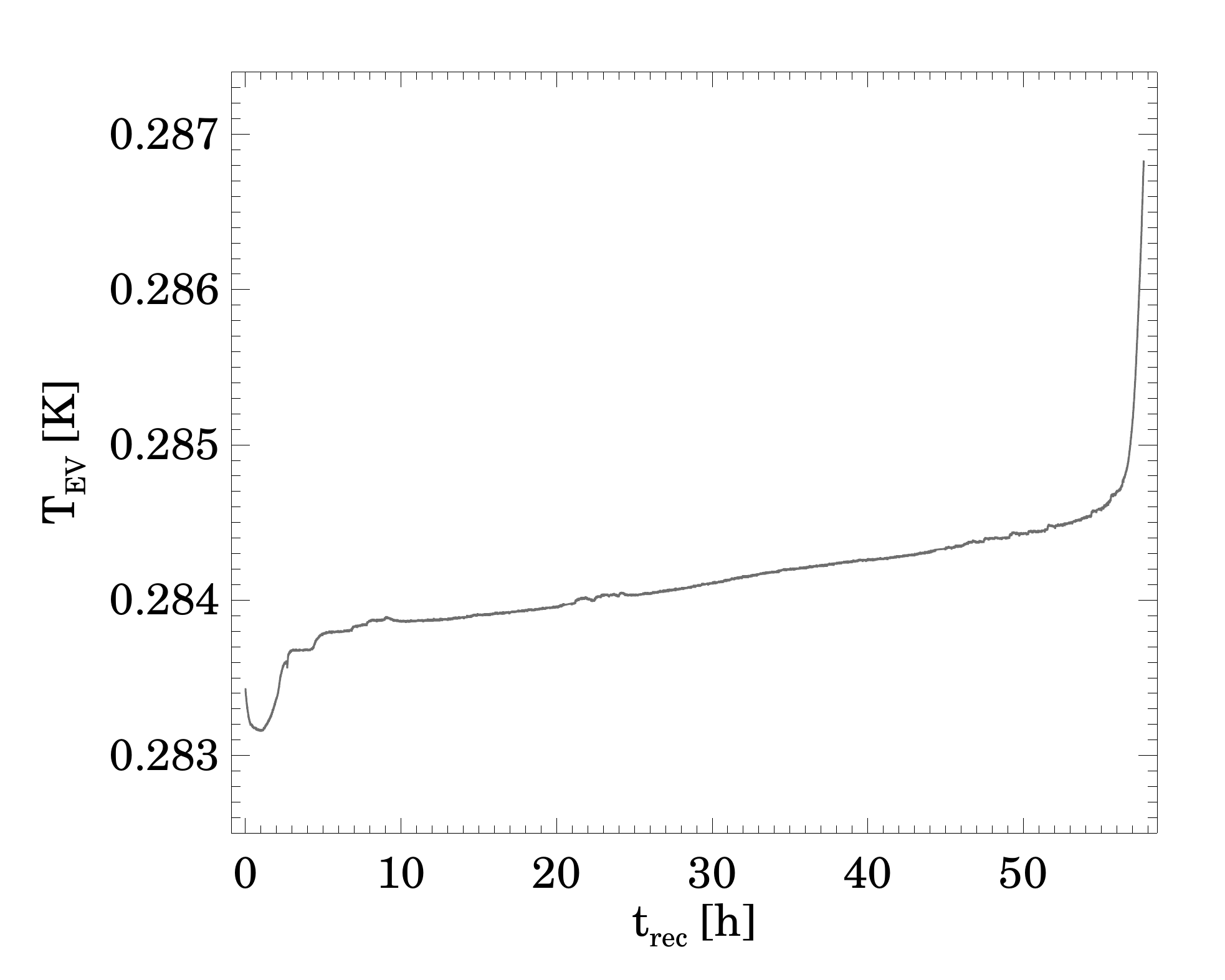}
\caption{{\sl Variation of evaporator temperature over a given cooler cycle (between OD917 and OD919.}
}
\label{coolingcycle}       % Give a unique label
\end{figure*}

During the mission a $^3$He cooler was used to ensure sub-Kelvin
temperature for the operation of PACS photometer. The
evaporation of $^3$He  provides a very stable temperature
environment at $\sim$300mK. After each cooler recycling
procedure, that takes about 2.5~h, there are about 2.5
ODs of PACS photometer observations possible.
The evaporator temperature ($T_{EV}$) rises towards end
of cooling cycles (Figure~\ref{coolingcycle}). 

\begin{figure*}[h!]
\centering
%\includegraphics[width=0.85\textwidth]{sdiff_TEV_multi.eps}
%\caption{{\sl Averaged differential CB signals as a function of evaporator temperature for measurements 
%executed with blue (at 70$\mu$m, upper panel), green (at 100$\mu$m, mid panel), 
%and red (at 160$\mu$m, lower panel) filter.   
%The observed trend can be well fitted by a linear (displayed as blue, green, and red lines).  
\includegraphics[width=0.85\textwidth]{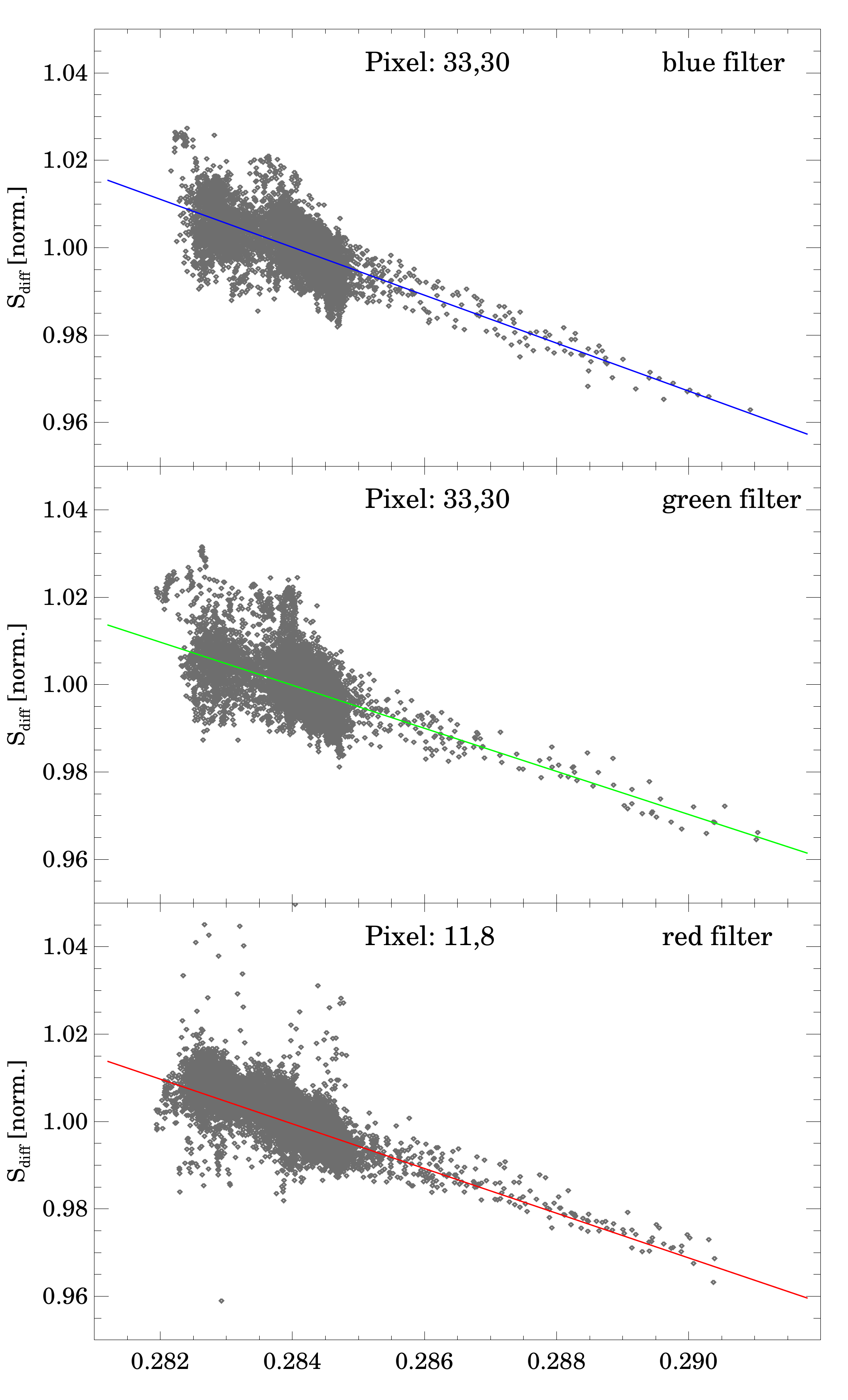}
\caption{{\sl Normalized differential CB signals of a selected pixel as a function of evaporator temperature for measurements 
executed with blue (at 70$\mu$m, upper panel), green (at 100$\mu$m, mid panel), 
and red (at 160$\mu$m, lower panel) filter.   
The observed trend can be well fitted by a linear (displayed as blue, green, and red lines).
 }
}
\label{tev}       % Give a unique label
\end{figure*}

By investigating the behaviour of $S_{diff}$
as a function of $T_{EV}$ (see Fig.~\ref{tev}) we found a clear correlation
between the two parameters for both the blue and the
red arrays. The observed trend can be well fitted by a
linear relationship enabling a correction for this effect.
It is important to note that measured flux densities of standard stars extracted
from PACS calibration observations showed very similar
trend with evaporator temperature and their 
photometry can be improved by
applying our pixel-based $T_{EV}$ correction for their
observations (Balog et al., this issue).

The correlation with $T_{EV}$ is very clear, but the related
corrections are typically well below 1\%, however, they vary slightly from pixel-to-pixel.
It is only $\sim$1.2\% of all
measurements for which this correction is larger
than 1\%!

\subsection{Correlation with temperature of focal plane unit}

After the $T_{EV}$ correction, we also looked for remaining trends and
correlations. We found that differential CS signals also depend on the FPU (Focal Plane
Unit) temperature ($T_{FPU}$). The FPU temperature ("BOL\_TEMP\_FPU\_ST" parameter 
in the housekeeping data) was  
measured at the 2K structure between two focal planes (M\"uller, 2009).
%\footnote{http://pacs.ster.kuleuven.be/Documents/WBS/FlightReports/012/001/tb\_phot\_v20
%090703.pdf} ).
%measured at the FPU structure close to the spacecraft
%L1 level as given in the instrument HK).
\begin{figure*}[h!!!!]
\centering
%\includegraphics[width=0.85\textwidth]{sdiff_fpu_multi.eps}
%\caption{{\sl Averaged differential CB signals as a function of FPU temperature (for blue, green, and red filter 
%observations). The plotted CB signals have been already corrected for $T_{EV}$ effect.
%The observed trend can be well fitted by a linear (displayed as blue, green, and red lines).  
\includegraphics[width=0.85\textwidth]{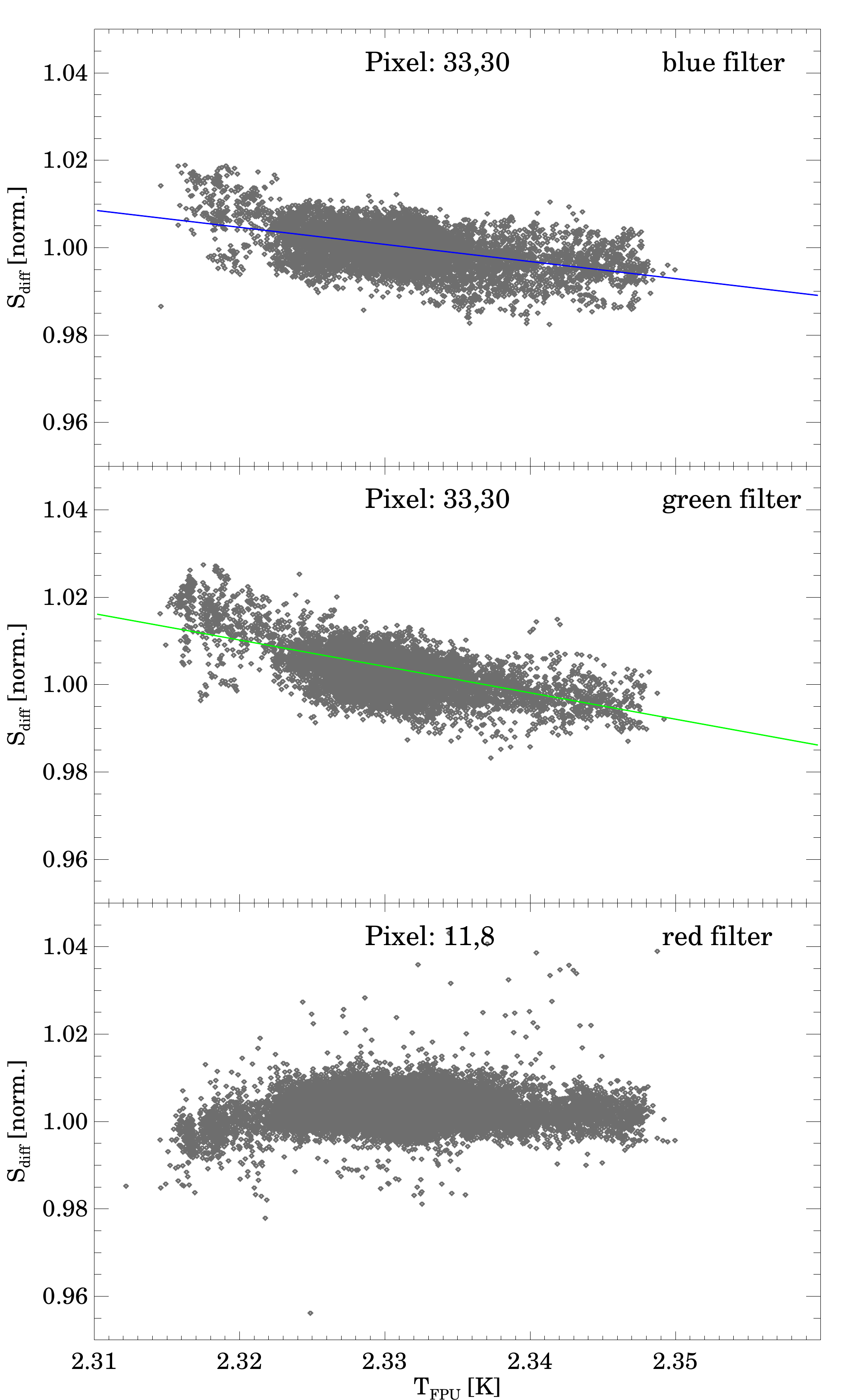}
\caption{{\sl Normalized differential CB signals for a selected pixel as a function of FPU temperature 
(for blue, green, and red filter 
observations). The plotted CB signals have already been corrected for $T_{EV}$ effect.
The observed trend can be well fitted by a linear (displayed as blue and green lines). 
 }
}
\label{fpu}       % Give a unique label
\end{figure*}

The observed trend is not as strong as in the case of $T_{EV}$ but can 
also be fitted by a linear relationship (Figure~\ref{fpu}). 
Interestingly, the red array
is not affected by this phenomenon. 
%Note that the $T_{FPU}$ values are 
%well correlated with the temperature measured at the outer part of CS1 integrating sphere, thus 
%the revealed trend is
Though the reason behind the observed trend has not been completely understood yet, it might be that the heat
   load of the FPU structure influences the bolometers via
   the Kevlar wires which connect the 300\,mK bolometers to
   the 2\,K FPU structure.
The variation of FPU temperature may affect calibration blocks only and 
 may not cause changes in the science observations.

\subsection{Drifting effect at early recycling times}
\begin{figure*}[h!!!!]
\centering
%\includegraphics[width=0.85\textwidth]{sdiff_trec_multi.eps}
%\caption{{\sl Normalized differential CB signals as a function of recycling time (for blue, green, and red filter 
%observations). The plotted CB signals have been already corrected for $T_{EV}$ effect.
%Model fits are shown as blue, green, and red lines.
\includegraphics[width=0.85\textwidth]{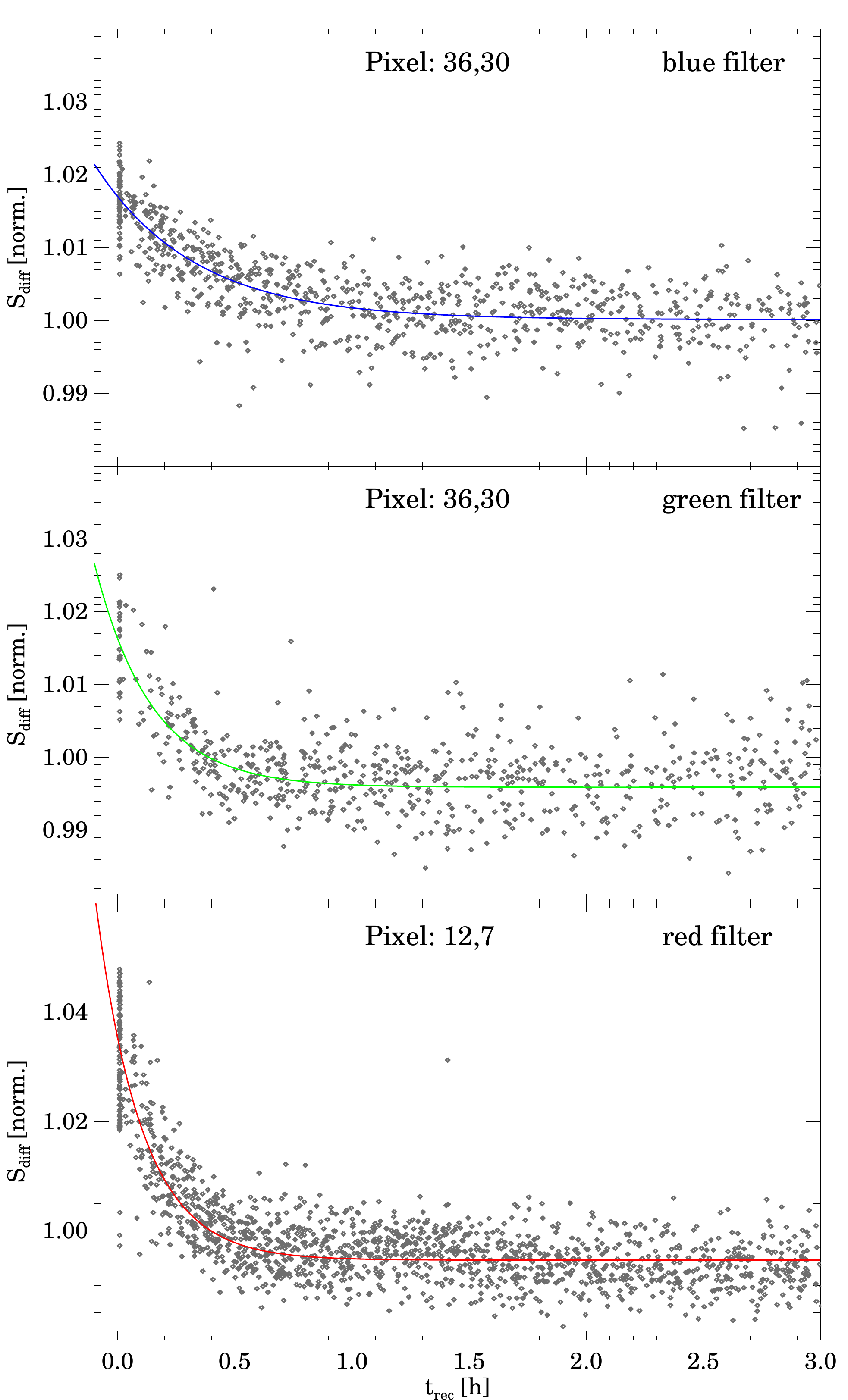}
\caption{{\sl Normalized differential CB signals for a selected pixel as a function of 
recycling time (for blue, green, and red filter 
observations). The plotted CB signals have already been corrected for $T_{EV}$ effect.
Model fits are shown as blue, green and red lines.
   }
}
\label{trec}       % Give a unique label
\end{figure*}

As Figure~\ref{trec} demonstrates, differential CS signals also depend on the time elapsed since the end of the
last cooling recycling procedure ($t_{rec}$). 
At the very beginning of PACS
observing sessions a stabilization effect can be recognized. 
This drift is only a very small effect and is only seen during the first
$\sim$30 minutes after the recycling has finished. In most cases this drift may
usually be covered in the initial slew to the first science target after the
recycling. The observed trend can be well fitted by a simple model: 
$$S_{diff}^{i,j} = a_0^{i,j} {\exp{(-t_{rec}/a_1^{i,j})} + a_2^{i,j}},$$
where $a_0^{i,j}, a_1^{i,j},$ and $a_2^{i,j}$ are parameters to be fitted.

\subsection{Evolution of differential CB signals}

\begin{figure*}[h!!!!!]
  \includegraphics[width=0.50\textwidth]{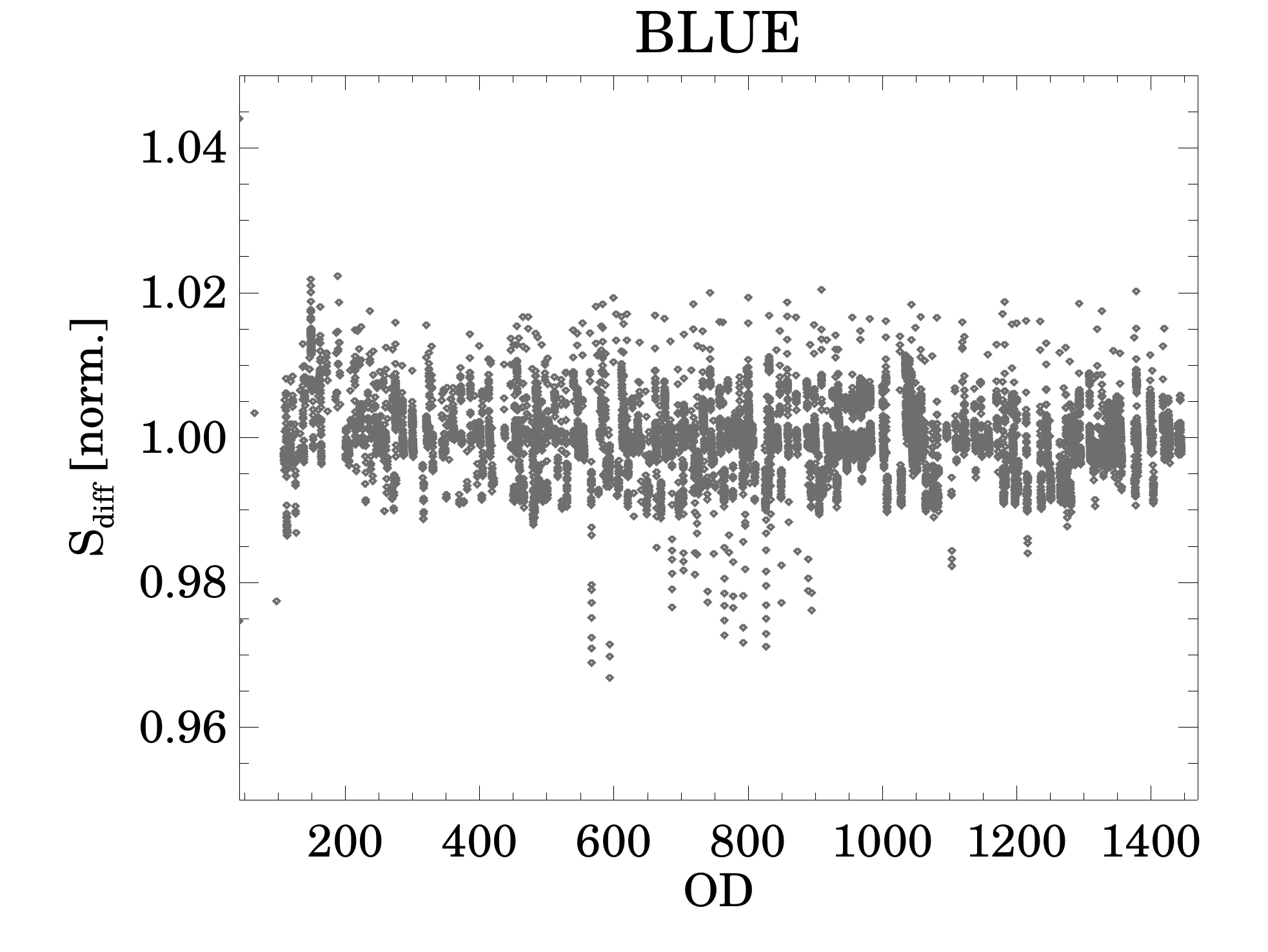}
  \includegraphics[width=0.50\textwidth]{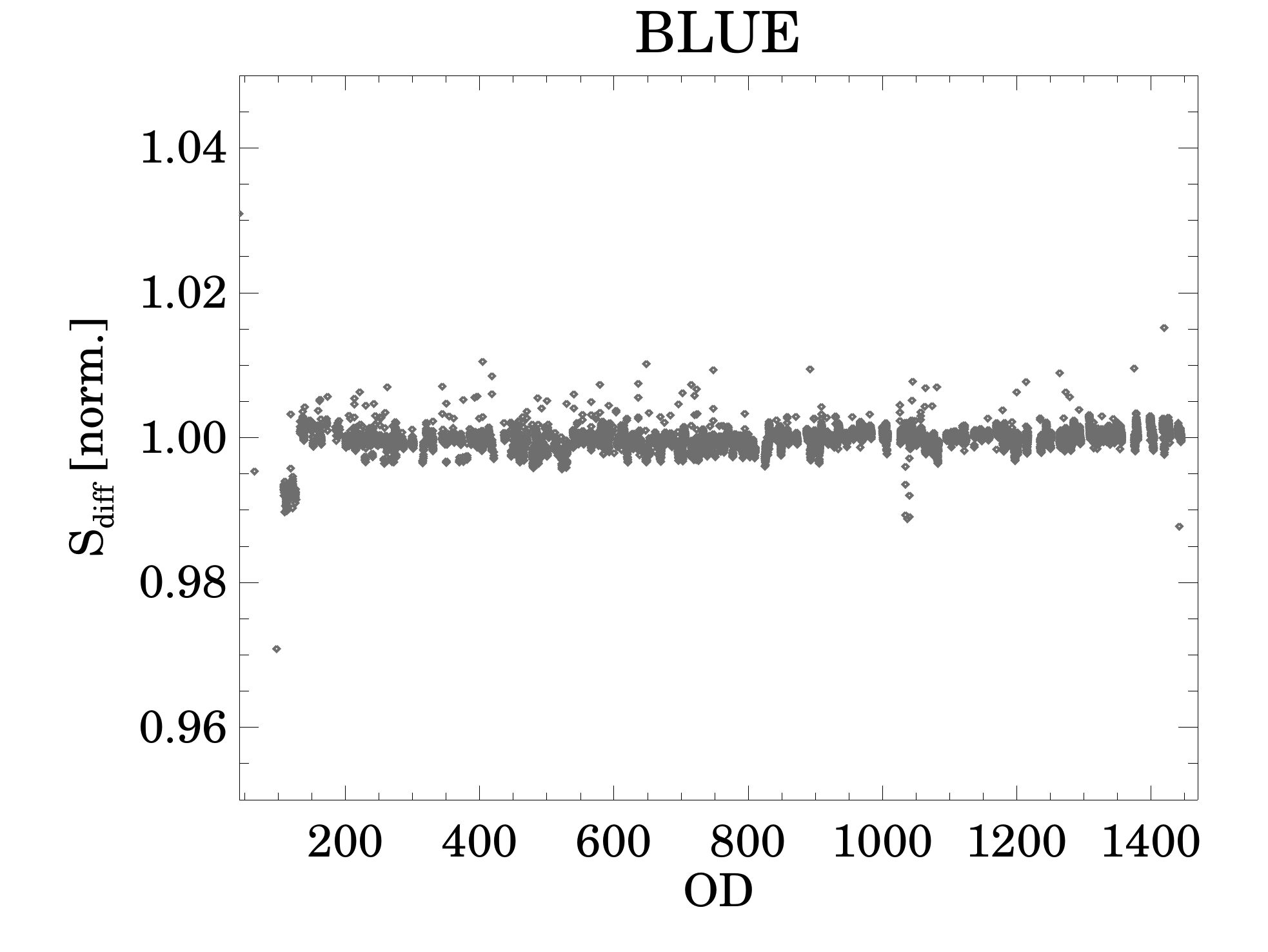}
  \includegraphics[width=0.50\textwidth]{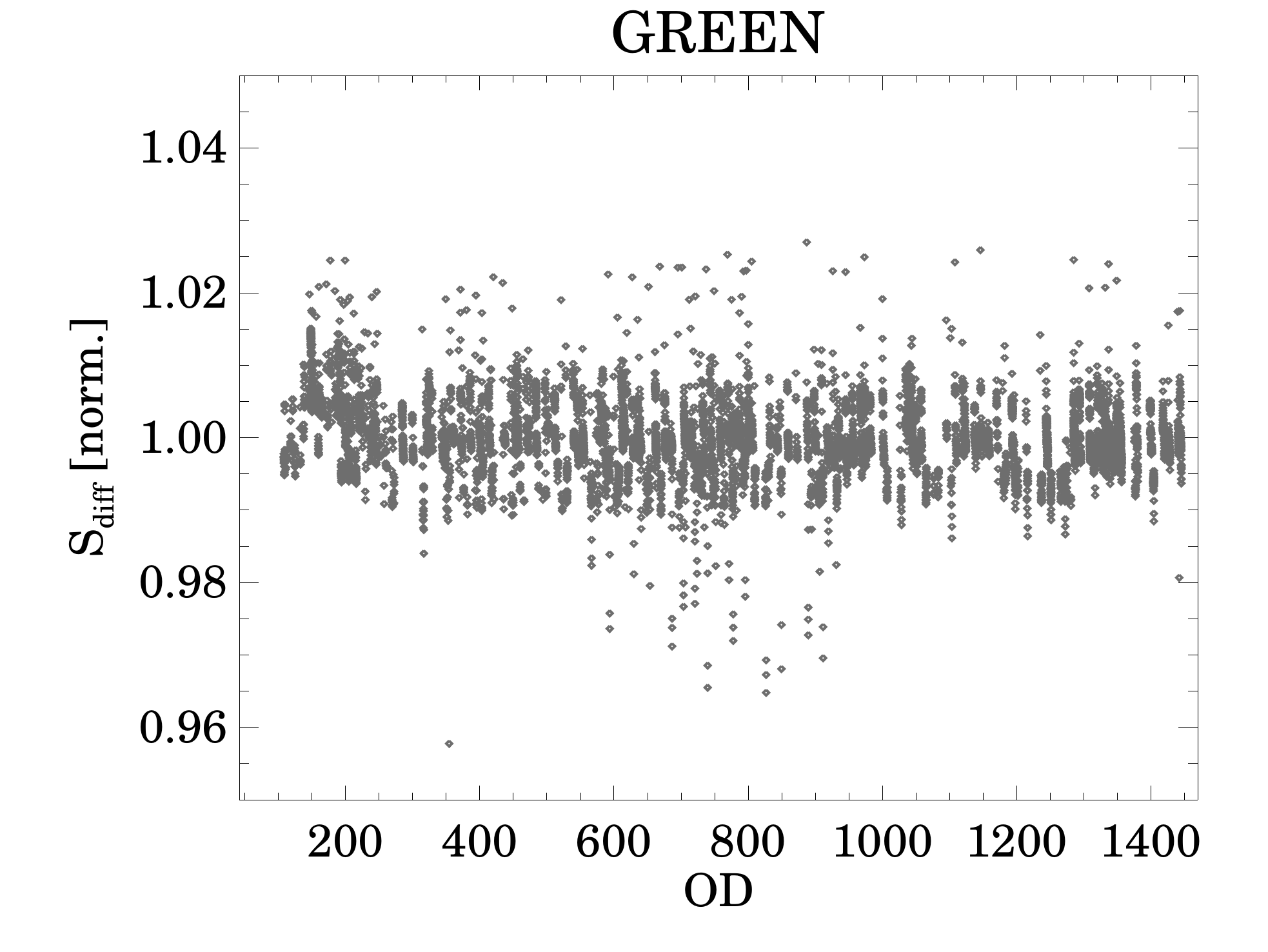}
  \includegraphics[width=0.50\textwidth]{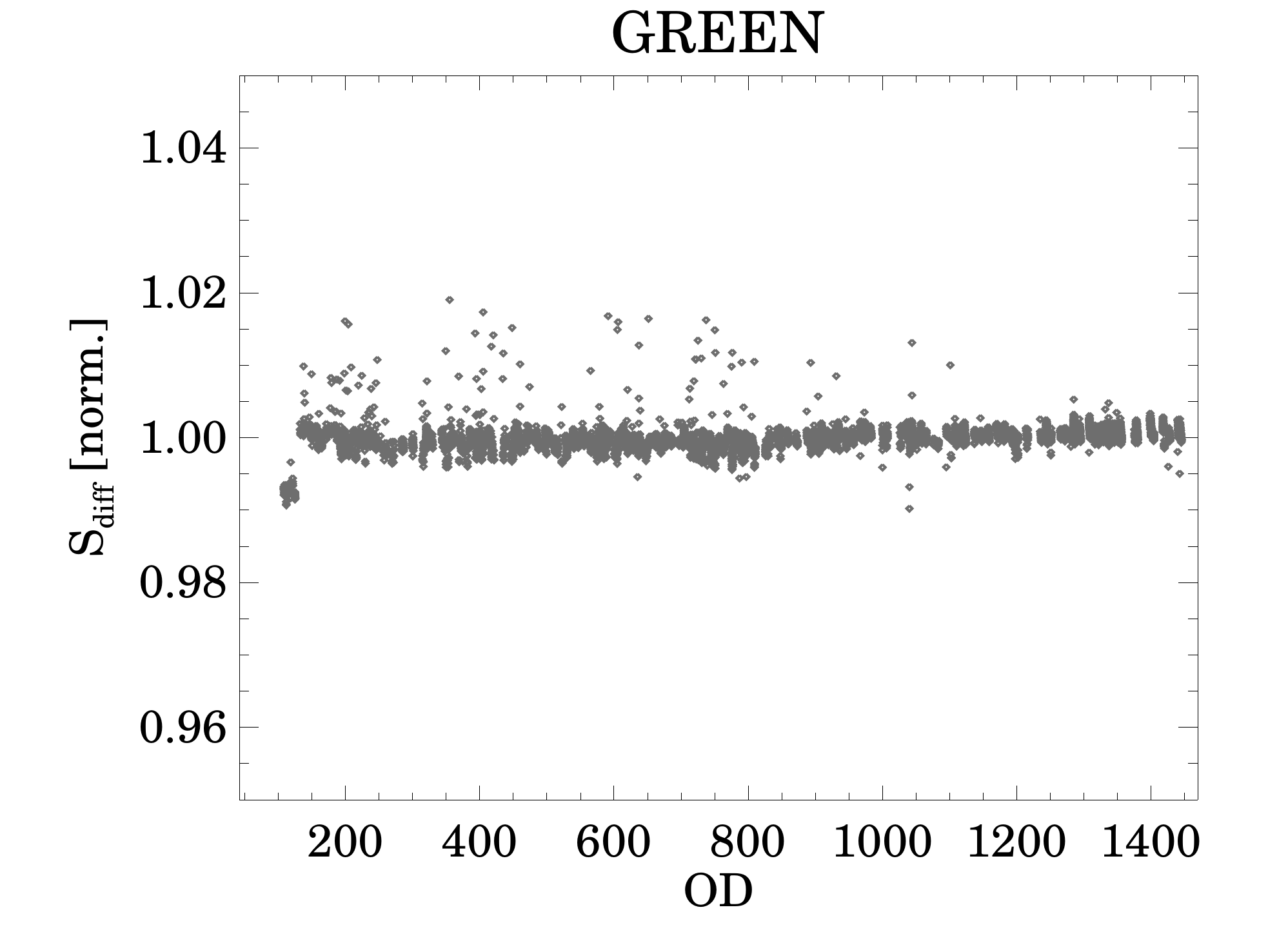}
  \includegraphics[width=0.50\textwidth]{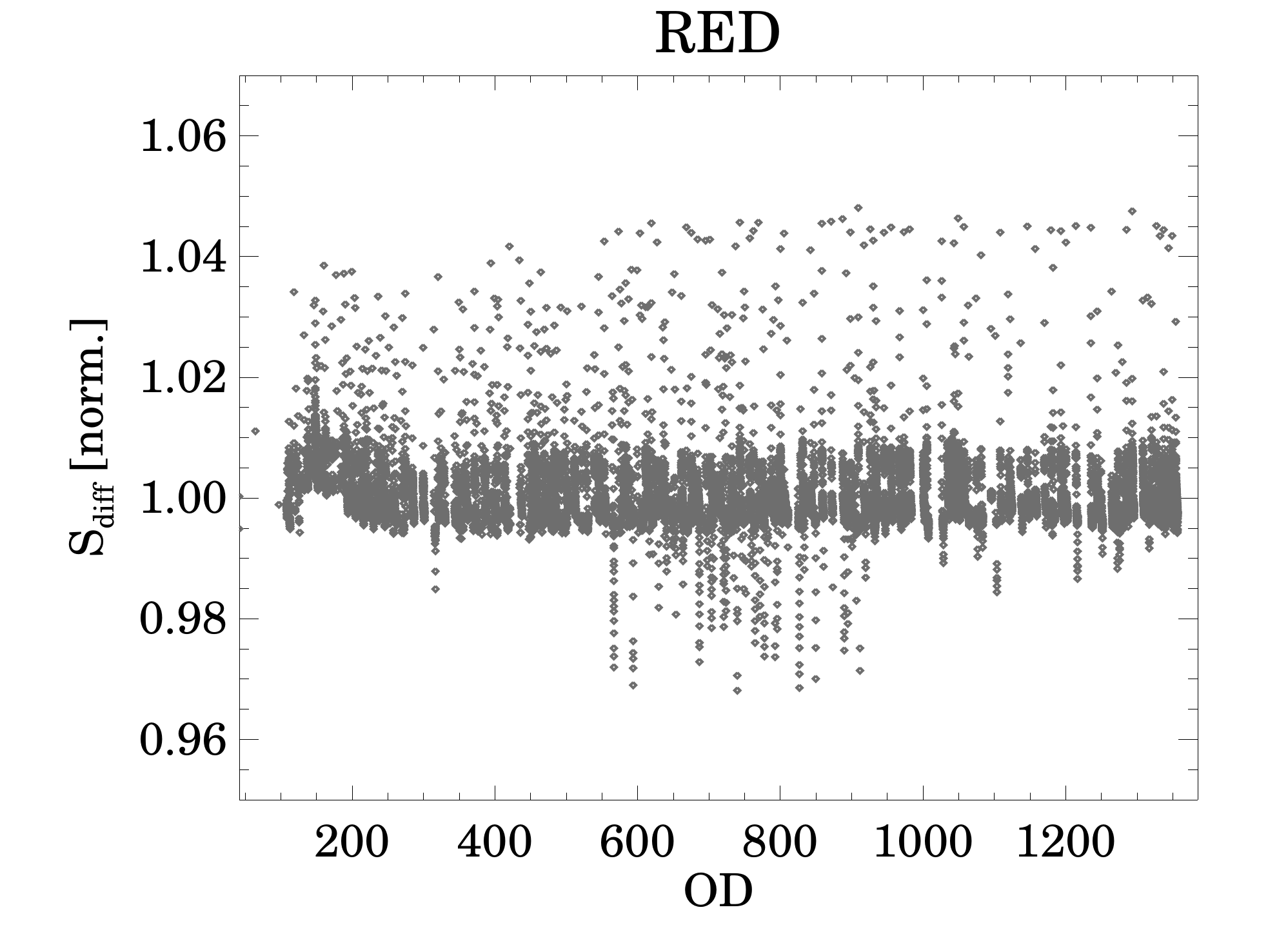}
  \includegraphics[width=0.50\textwidth]{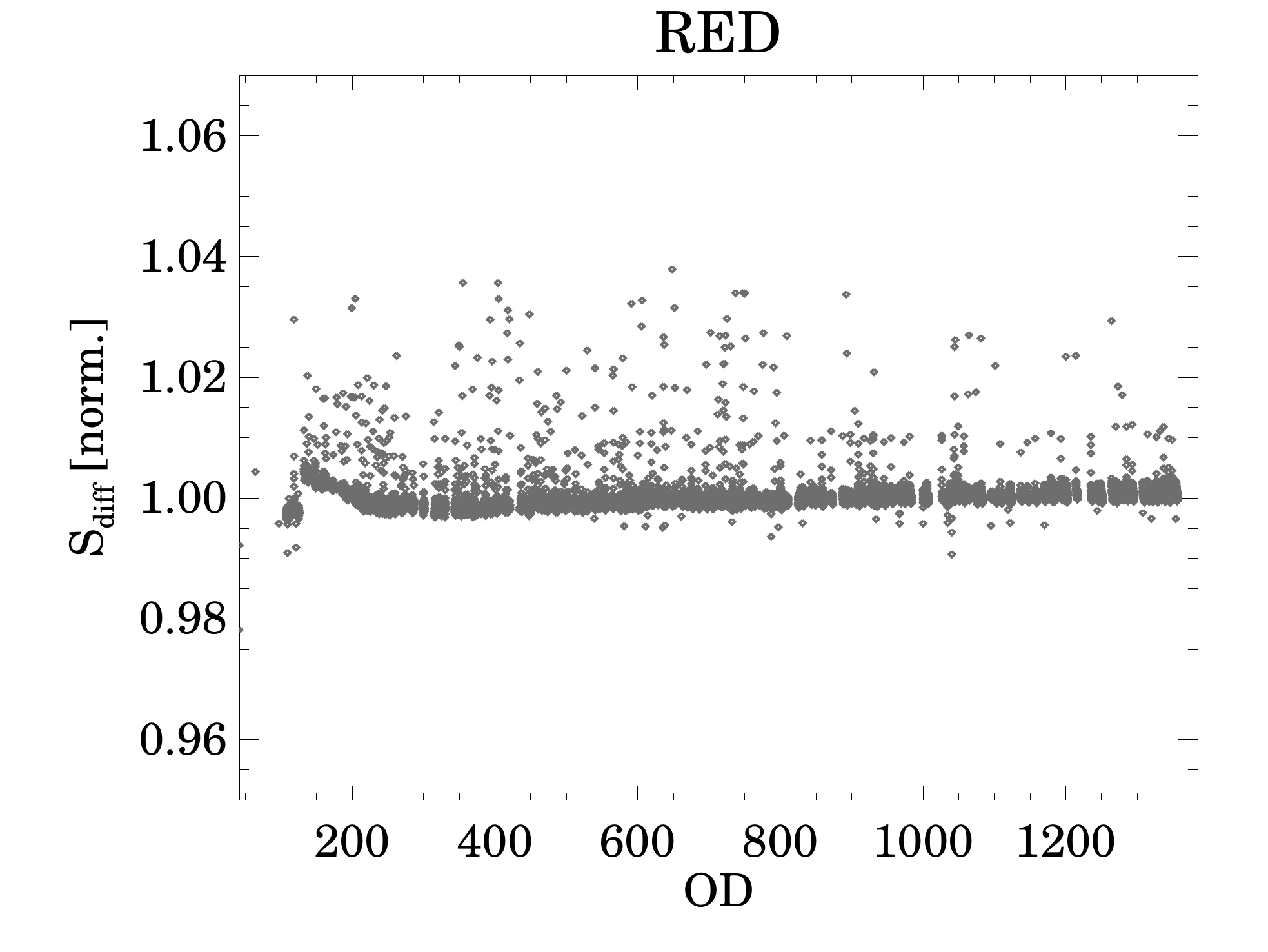} 
  %\includegraphics[width=0.50\textwidth]{blue_l1-eps-converted-to.pdf}
  %\includegraphics[width=0.50\textwidth]{blue_l4-eps-converted-to.pdf}
  %\includegraphics[width=0.50\textwidth]{green_l1-eps-converted-to.pdf}
  %\includegraphics[width=0.50\textwidth]{green_l4-eps-converted-to.pdf}
  %\includegraphics[width=0.50\textwidth]{red_l1-eps-converted-to.pdf}
  %\includegraphics[width=0.50\textwidth]{red_l4-eps-converted-to.pdf} 
\caption{{\sl Left panels: Averaged differential CS signals as a function of operational day
without any correction for 
blue (upper panel), green (mid panel), and red filter (lower panel) CB measurements. 
Right panels: Averaged differential CS signals as a function of operational day 
after corrections for $T_{EV}$, $T_{FPU}$ and $t_{rec}$ were applied.
The red PACS detector consisted of 2 bolometer matrices (matrix\,9 and matrix\,10).
After OD1375 one of the two red channel subarrays (matrix\,9) was affected by a serious anomaly.
Therefore red data were not plotted for OD $>$1375.
%taken with the PACS Photometer from OD1375
%Because of the operational issue with one subarray of the red detector 
}
}
\label{signalevolution}       % Give a unique label
\end{figure*}

As Figure~\ref{signalevolution} shows, the level of differential signals and thus 
 the PACS bolometer response was stable over the entire 
mission lifetime. 
The stability of the differential CS signal even without any correction 
was less than $\sim$0.5\% (standard deviation) or $\sim$8\% (peak-to-peak)
in blue/green/red observations (Fig.~\ref{signalevolution}, left panels or in 
Table\ref{tab:1}).
By applying all of our corrections we can achieve a stability level of 
$\sim$0.12\% for blue array observations and 0.2\% (standard deviation) for data obtained using 
the red bolometer (Fig.~\ref{signalevolution}, right panels). 
The peak-to-peak variation of the final data set is lower than 
2, 3 and 5\% in blue/green/red CB observations. 
As Table~\ref{tab:1} shows the correction related to $T_{EV}$ is by far the most important 
among all corrections we applied.
\begin{table}[h!]
\caption{Influence of corrections on the standard deviation of differential CS signals}
\label{tab:1}       % Give a unique label
\begin{tabular}{lccc}
\hline\noalign{\smallskip}
Processing  & \multicolumn{3}{|c|}{Standard deviation of differential CS signals} \\
  stage     & \multicolumn{3}{|c|}{averaged for bolometer arrays [\%]} \\ 
   & Blue filter & Green filter & Red filter \\
\noalign{\smallskip}\hline\noalign{\smallskip}
Before corrections were applied & 0.49 & 0.48  & 0.53 \\
After correction for $T_{EV}$ was applied & 0.22 & 0.22  & 0.34 \\
After $T_{EV}$ and $t_{rec}$ corrections & 0.17 & 0.16  & 0.18 \\
After all corrections were applied & 0.12 & 0.13  & 0.18 \\
\noalign{\smallskip}\hline
\end{tabular}
\end{table}
Due to this excellent stability the calibration of scientific measurements can be done 
without using the PACS CBs directly, only the $T_{EV}$
   correction -- which will be part of future HIPE \& pipeline versions --
   is based on information extracted from the frequent measurements
   of the PACS internal calibration sources.

The significantly decreased scatter of differential signals allowed us to 
 search for remaining weaker trends in the data set.
Our most interesting findings are as follow:
\begin{itemize}
\item PACS photometer was operating in dual band mode taking data with a red filter and a blue or green filter 
simultaneously on two bolometer arrays. We found that the filter 
of the blue array has an effect on the signal level measured in the red band. 
As Figure~\ref{filterselection} demonstrates, the differential red
   signals are higher when the 70$\mu$m filter was placed in the
   blue channel and lower when the 100$\mu$m filter was used.
   This could be caused by slightly different straylight/reflection
   patterns inside PACS which change with filter wheel positions.
Although the difference is very small, a two-sided Kolmogorov-Smirnov test shows that the two distributions 
are significantly different.  
\begin{figure*}
  \includegraphics[width=0.50\textwidth]{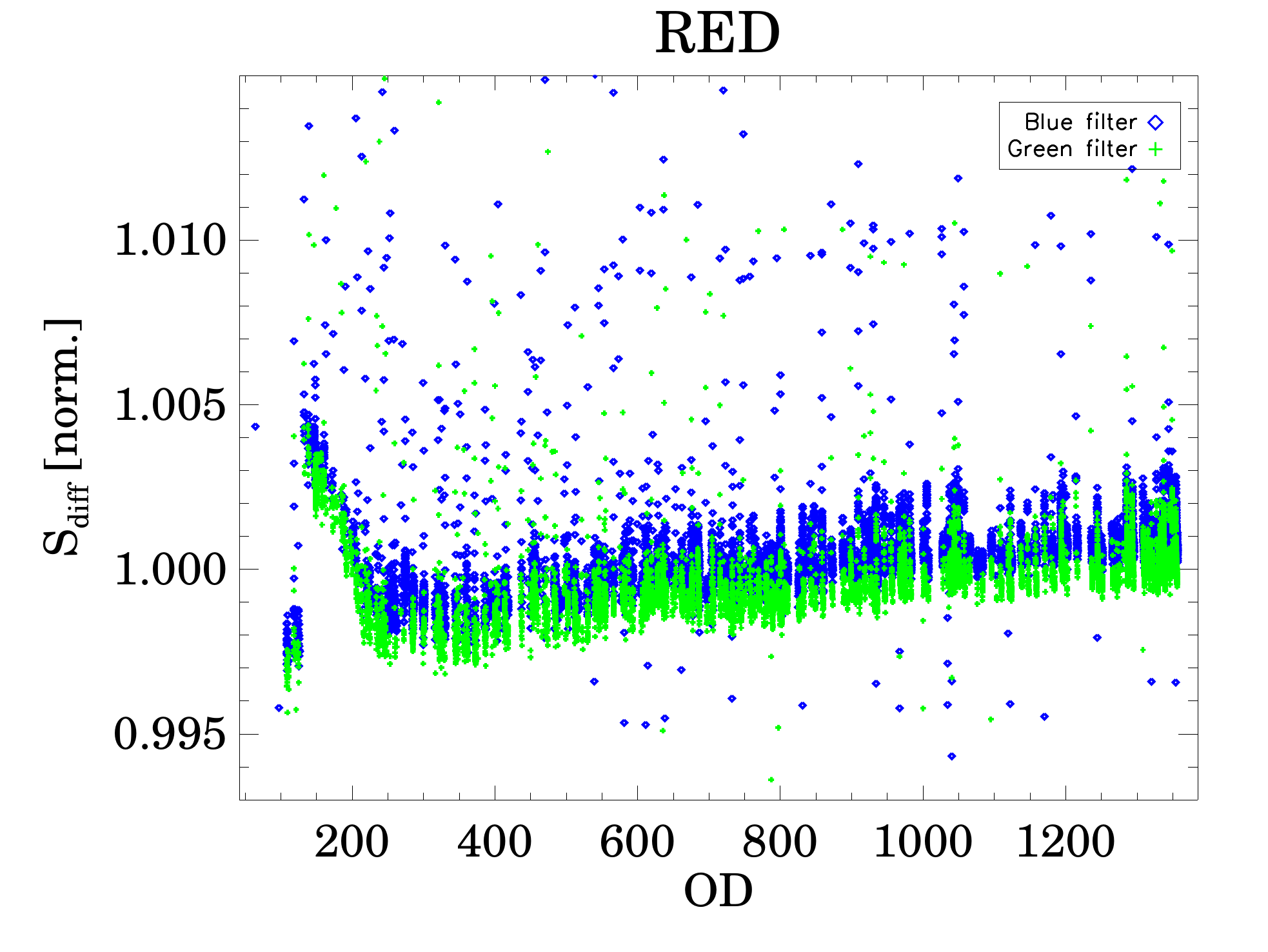}
  \includegraphics[width=0.50\textwidth]{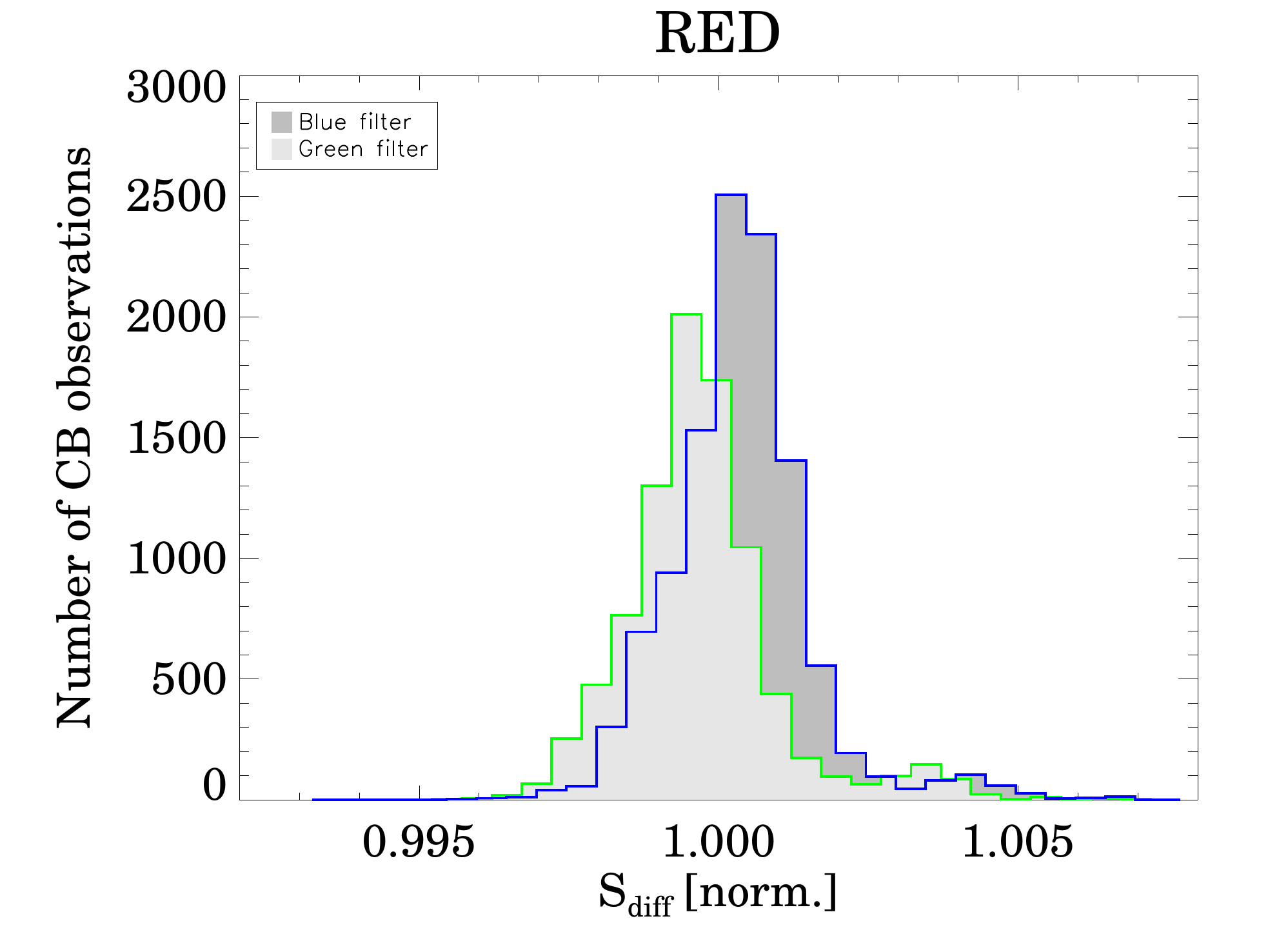}
  %\includegraphics[width=0.50\textwidth]{red_l4_filter-eps-converted-to.pdf}
  %\includegraphics[width=0.50\textwidth]{red_blue_green_histo-eps-converted-to.pdf}
% figure caption is below the figurered_blue_green_histo.eps
\caption{{\sl Left panel: Averaged differential CS signals for the red array as a function of operational day. 
Red data points obtained with the blue or green short wavelength
PACS filter selection are displayed with different symbols. On the right panel we present the
distribution of these data.}
}
\label{filterselection}       % Give a unique label
\end{figure*}

\item  The red PACS detector is assembled from 2 bolometer matrices.
After OD1375 one of the two red channel subarrays (matrix\,9) was affected by a serious anomaly and 
therefore became unusable. As Figure~\ref{matrix} (left) shows the other subarray (matrix\,10) remained stable  
the measured signal level obtained with this part of the detector was not changed after this event. 
\item All 3 bands show an initial jump/stabilization phase during the
first $\sim$280 ODs (Figure~\ref{matrix}, right). The variation of CB signals is quite similar 
for all blue array observations (blue/green). The reason of this initial stabilization phase
is not understood, {but coincides with small
  variations in the detector setup (orbit prologue) during
  the early mission phase (see Performance Verification Plan,
  reference below).
There is an additional weak trend in the
  variation of red signals with operational days.
After OD 250
  the differential CB signals measured in the red band slowly
  increase with increasing OD (Fig.~\ref{filterselection} left).
Nielbock et al. (this issue) found similar effects in the
  chop-nod observations of the fiducial stars that they attribute
  to a change in the temperature gradient on the primary mirror.
  Though CSs do not see the primary mirror directly, their signals
  might be influenced via straylight.}
{We also note that during the commisioning phase of the PACS Photometer 
the detector bias voltages were modified at OD128 (for more details see Billot et al., 2010). 
The discontinuity seen in the differential signal levels at around this epoch in Figure~\ref{matrix} (right) 
might be due to this bias change.}

\begin{figure*}
  \includegraphics[width=0.50\textwidth]{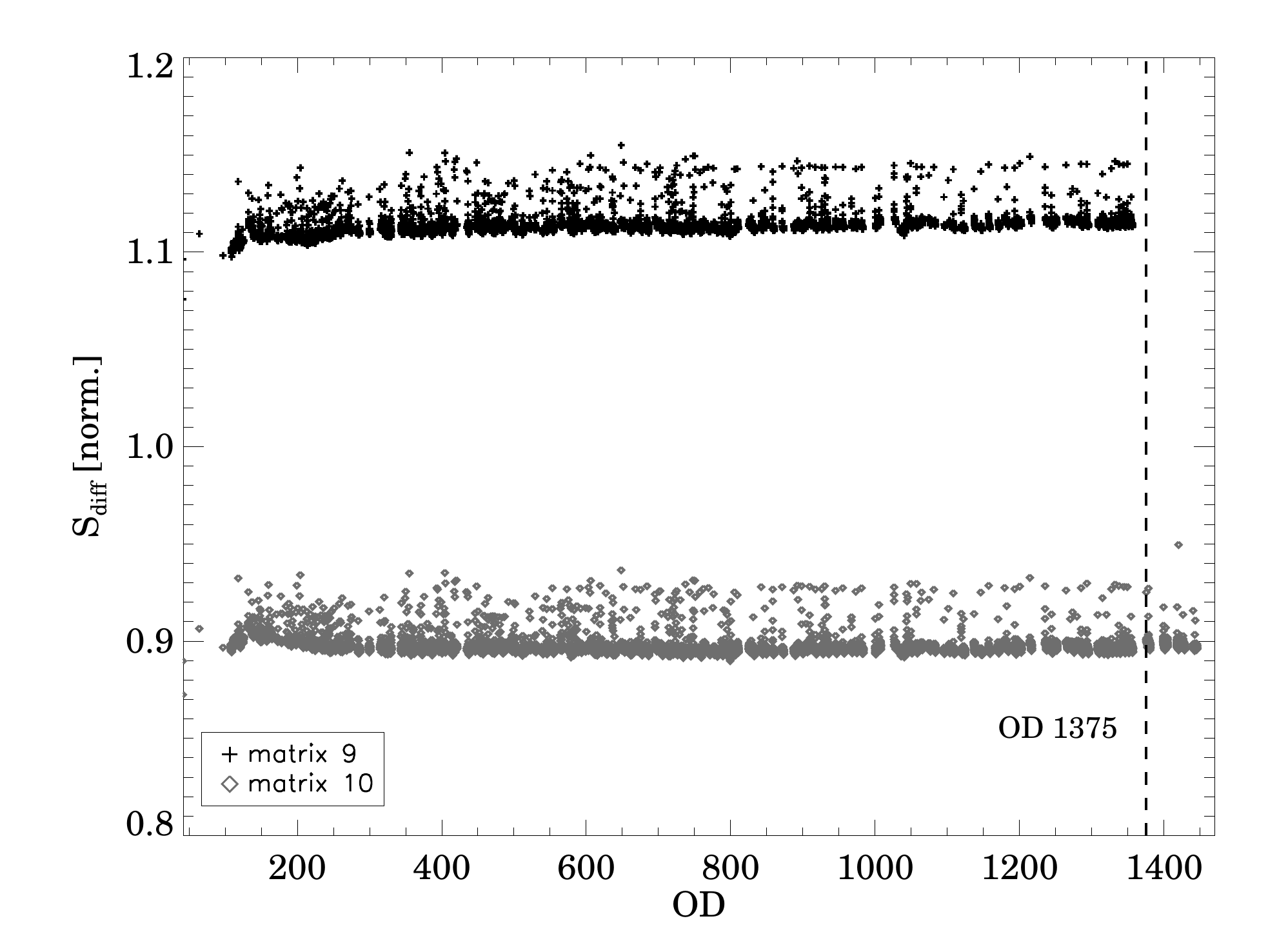}
  \includegraphics[width=0.50\textwidth]{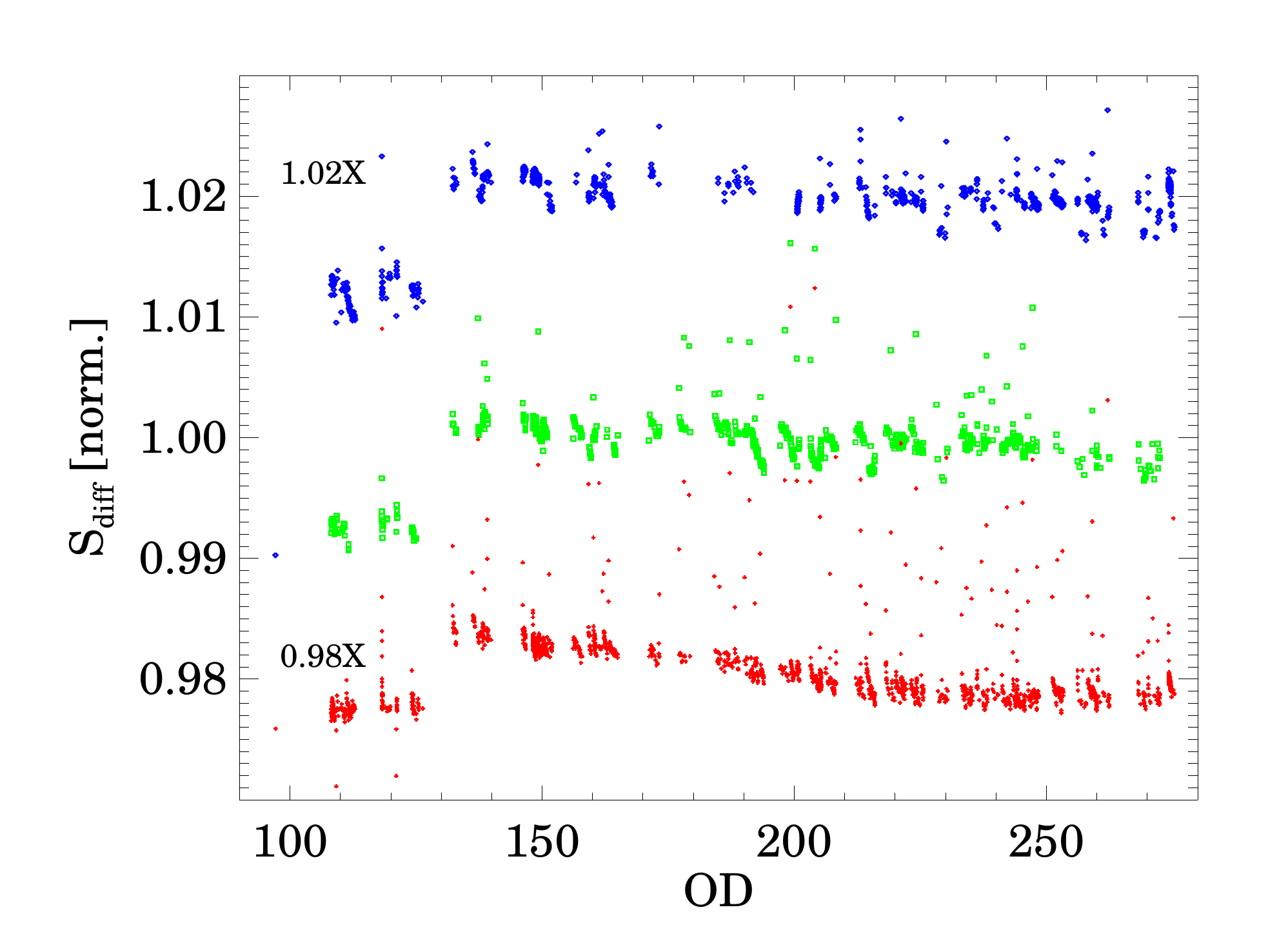}
  %\includegraphics[width=0.50\textwidth]{red_l4_filter-eps-converted-to.pdf}
  %\includegraphics[width=0.50\textwidth]{red_blue_green_histo-eps-converted-to.pdf}
% figure caption is below the figurered_blue_green_histo.eps
\caption{{\sl Left panel: Differential CB signals averaged for the two subarrays of the 
red detector (matrix 9 and matrix 10) as a function of OD. 
Right panel: Averaged differential CB signals (for blue, green, and red filter 
observations) as a function of OD for the early phase of the mission. 
Note that signals measured with the blue and red filters have been multiplied 
by a factor of 1.02 and 0.98, respectively, in order to avoid overlapping data points.}
}
\label{matrix}       % Give a unique label
\end{figure*}
\item {PACS instrument campaigns -- that took about  2.5 ODs -- were interrupted 
by {\sl Daily TeleCommunication
Periods} (DTCP) in which the previously collected data were downlinked to Earth. 
Before OD670 the PACS detector was put in safe mode during DTCP and the bolometers were not biased.
After that, however, the operational strategy was
changed and the PACS bolometers were kept biased
during DTCPs.
%Before OD676 the PACS detector were not biased during
%  these DTCPs. After that, however, the operational strategy was
%  changed, between OD 676 and OD 742 the usage of the PACS photometer
%  was restricted to 1.5 ODs and later on to 2.5 ODs without touching
%  the detector biases during DTCPs.
%was switched off during these DTCPs. 
%After that, however, the operational strategy was changed, between OD676 and OD742 
%the usage of PACS photometer was restricted to 1.5 ODs without switching the detector off, 
%while after OD 742 the detector was not switched off at all during 
% DTCPs.   
\begin{figure*}
  \includegraphics[width=0.50\textwidth]{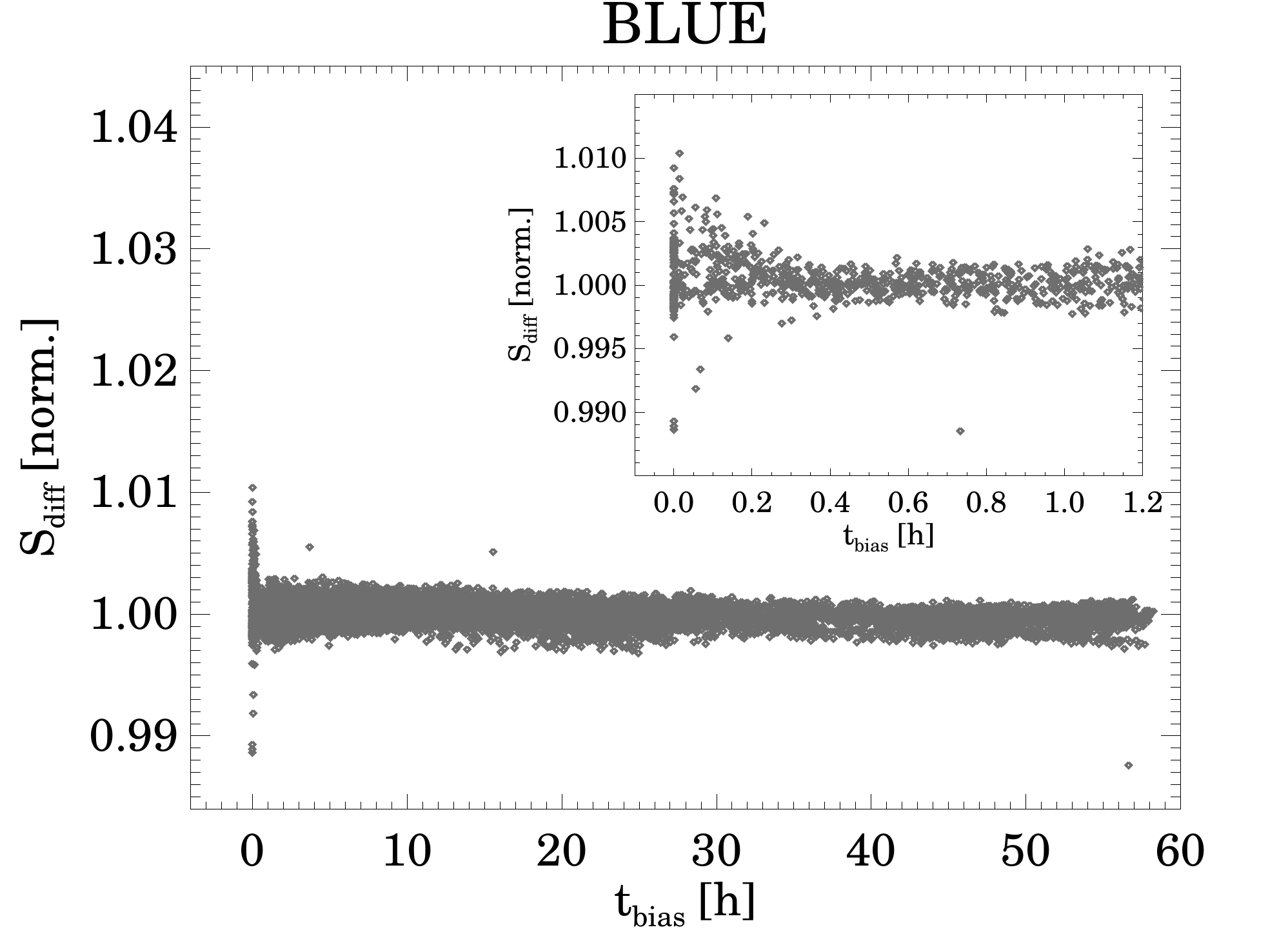}
  \includegraphics[width=0.50\textwidth]{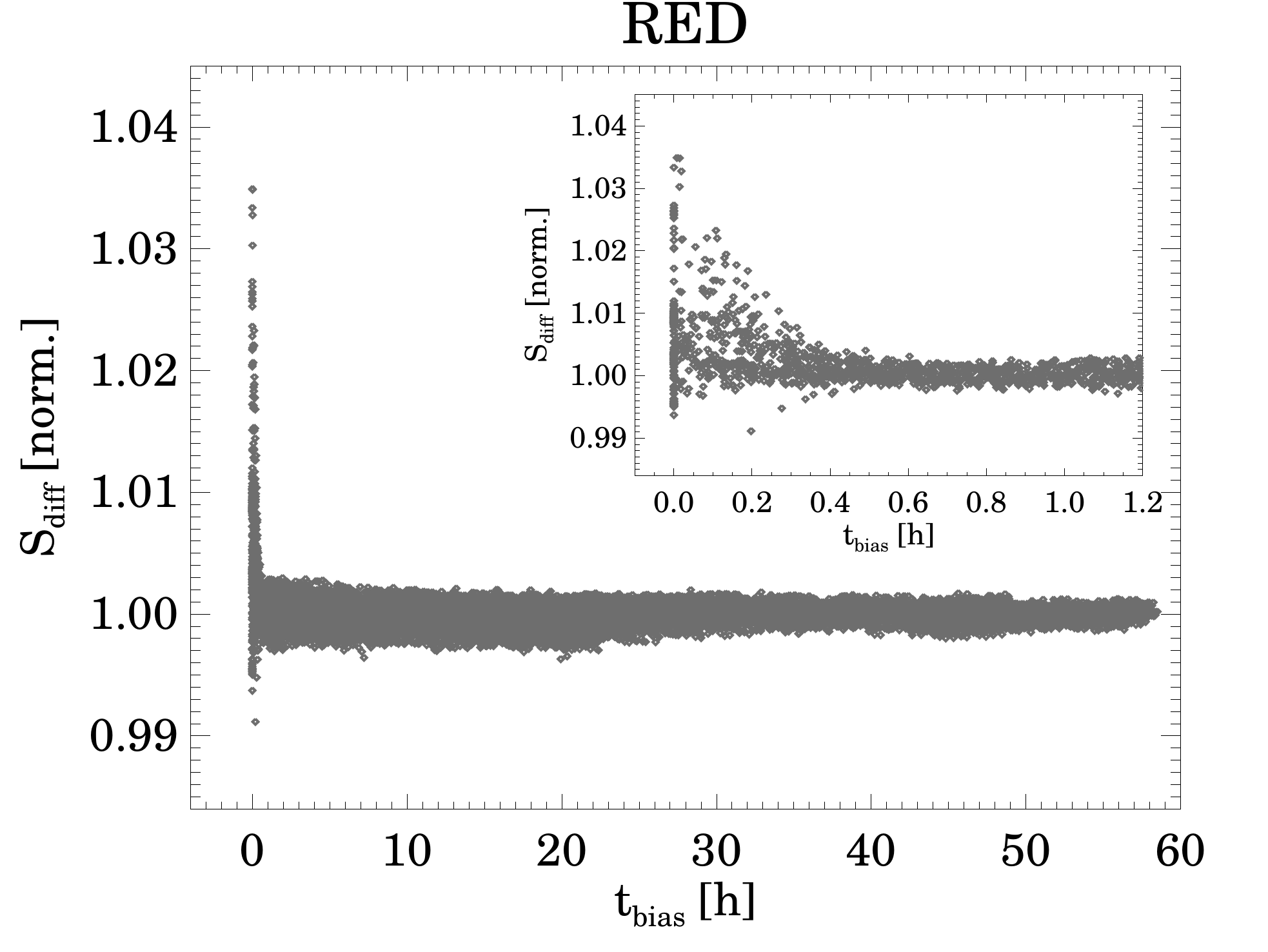}
  %\includegraphics[width=0.50\textwidth]{red_l4_filter-eps-converted-to.pdf}
  %\includegraphics[width=0.50\textwidth]{red_blue_green_histo-eps-converted-to.pdf}
% figure caption is below the figurered_blue_green_histo.eps
\caption{{\sl Averaged differential CB signals for blue and red filter 
observations as a function of $t_{bias}$.}
}
\label{tbias}       % Give a unique label
\end{figure*}
Figure \ref{tbias} shows the corrected 
differential CB signals as a function of $t_{bias}$
(the time elapsed since the last biasing of the detectors) for 
measurements executed with blue (left) and red filter (right), respectively.
Observations obtained in the early phase of the
mission ($<$ OD200) were discarded and not displayed in the figures
(in order to avoid early mission stabilization period, as described in the previous paragraph).
As these figures demonstrate, most of the remaining outliers are related to
observations performed very early after the preceding biasing, implying 
that these processes are also accompanied by a 
stabilization effect. }
\item We note that variation of solar aspect angle has no influence on signal levels.
\end{itemize}


\section{Summary}

Using our database of calibration block observations we investigated the long term behaviour of
differential CS signals which characterize the evolution of the bolometer response as well.
We revealed that variation of evaporator temperature and FPU temperature cause changes in
differential signals. Signal levels also show a short drift in the first ~0.5h of observing sessions after
the cooler recycling.
We developed correction methods for all three effects. By applying these corrections the standard
deviation of differential CS signals could be decreased significantly.
The main cause of differential signal changes is the variation of evaporator 
temperature. We note that measured
flux densities of standard stars show also a similar dependency on $T_{EV}$.
A task that performs this correction will be available in HIPE~12.0. 
%Our correction derived from
%the CB analysis will be included in the standard pipeline processing scripts after the validation
%process is completed. 
This will improve the photometry for measurements taken either directly after
the cooler recycling or very late towards the end of the cooler hold time.
The drifting effect observed at low recycling time is more important for the red array than for the blue.
%The TEV and the trec correction could be relevant for variability programmes.
The bolometer response turned out to be very stable over the entire mission lifetime 
thus the individual calibration blocks do not need to be used in the processing
  of the corresponding science data.
No aging effect or degradation of the photometric system during the mission 
lifetime has been found.

\begin{acknowledgements}
We thank our anonymous referee whose comments significantly 
improved the manuscript.
This research has been supported by the PECS programme 
of the Hungarian Space Office and the European Space Agency 
(contact number: 98073). CK and AM acknowledge the support
of the Hungarian Research Fund (OTKA K-104607) and that of 
the Bolyai Research Fellowship of the Hungarian Academy of Sciences.  
\end{acknowledgements}



\end{document}